\documentclass[aps,pra,twocolumn,superscriptaddress]{revtex4-2}
\usepackage{amsmath,amssymb,amsfonts}
\usepackage{graphicx}
\usepackage{textcomp}

\usepackage{braket}
\usepackage{subfigure}
\usepackage{float}
\usepackage{caption}
\usepackage{comment}
\usepackage{algorithm}
\usepackage{hyperref}
\usepackage{makecell}
\usepackage[noend]{algpseudocode}

\usepackage{enumerate}
\usepackage{array}
\usepackage{xcolor}

\usepackage{bm}
\usepackage{booktabs}

\newcommand{\removelatexerror}{\let\@latex@error\@gobble}

\makeatletter
\newenvironment{breakablealgorithm}
  {% \begin{breakablealgorithm}
   \begin{center}
     \refstepcounter{algorithm}% New algorithm
     \hrule height.8pt depth0pt \kern2pt% \@fs@pre for \@fs@ruled 画线
     \renewcommand{\caption}[2][\relax]{% Make a new \caption
       {\raggedright\textbf{\ALG@name~\thealgorithm} ##2\par}%
       \ifx\relax##1\relax % #1 is \relax
         \addcontentsline{loa}{algorithm}{\protect\numberline{\thealgorithm}##2}%
       \else % #1 is not \relax
         \addcontentsline{loa}{algorithm}{\protect\numberline{\thealgorithm}##1}%
       \fi
       \kern2pt\hrule\kern2pt
     }
  }{% \end{breakablealgorithm}
     \kern2pt\hrule\relax% \@fs@post for \@fs@ruled 画线
   \end{center}
  }
\makeatother

\def\BibTeX{{\rm B\kern-.05em{\sc i\kern-.025em b}\kern-.08em
    T\kern-.1667em\lower.7ex\hbox{E}\kern-.125emX}}

\begin{document}

%\preprint{}

\title{Robust Optimization for Quantum Reinforcement Learning Control using Partial Observations}


\author{Chen Jiang}
\author{Yu Pan}
\email[]{ypan@zju.edu.cn}
\author{Zheng-Guang Wu}
\affiliation{State Key Laboratory of Industrial Control Technology, Institute of Cyber-Systems and Control, College of Control Science and Engineering, Zhejiang University, Hangzhou, 310027, China}

\author{Qing~Gao}
\affiliation{The School of Automation Science and Electrical Engineering and Beijing Advanced Innovation Center for Big Data and Brain Computing, Beihang University, Beijing 100191, China}

\author{Daoyi~Dong}
\affiliation{The School of Engineering and Information Technology, University of New South Wales, Canberra, ACT 2600, Australia}

%\thanks{}
\thanks{This work was supported by the National Natural Science Foundation of China (No. 62173296), the Alexander von Humboldt Foundation, Germany and in part by the Australian Research Council’s Discovery Projects Funding Scheme under Project DP190101566.}


\date{\today}

\begin{abstract}
The current quantum reinforcement learning control models often assume that the quantum states are known a priori for control optimization. However, full observation of quantum state is experimentally infeasible due to the exponential scaling of the number of required quantum measurements on the number of qubits. In this paper, we investigate a robust reinforcement learning method using partial observations to overcome this difficulty. This control scheme is compatible with near-term quantum devices, where the noise is prevalent and predetermining the dynamics of quantum state is practically impossible. We show that this simplified control scheme can achieve similar or even better performance when compared to the conventional methods relying on full observation. We demonstrate the effectiveness of this scheme on examples of quantum state control and quantum approximate optimization algorithm. It has been shown that high-fidelity state control can be achieved even if the noise amplitude is at the same level as the control amplitude. Besides, an acceptable level of optimization accuracy can be achieved for QAOA with noisy control Hamiltonian. This robust control optimization model can be trained to compensate the uncertainties in practical quantum computing.
\end{abstract}

%\keywords{}

\maketitle

\section{Introduction}
\label{sec:introduction}
\begin{figure*}[htbp]
\centerline{\includegraphics[height=5.39cm,width=15.05cm]{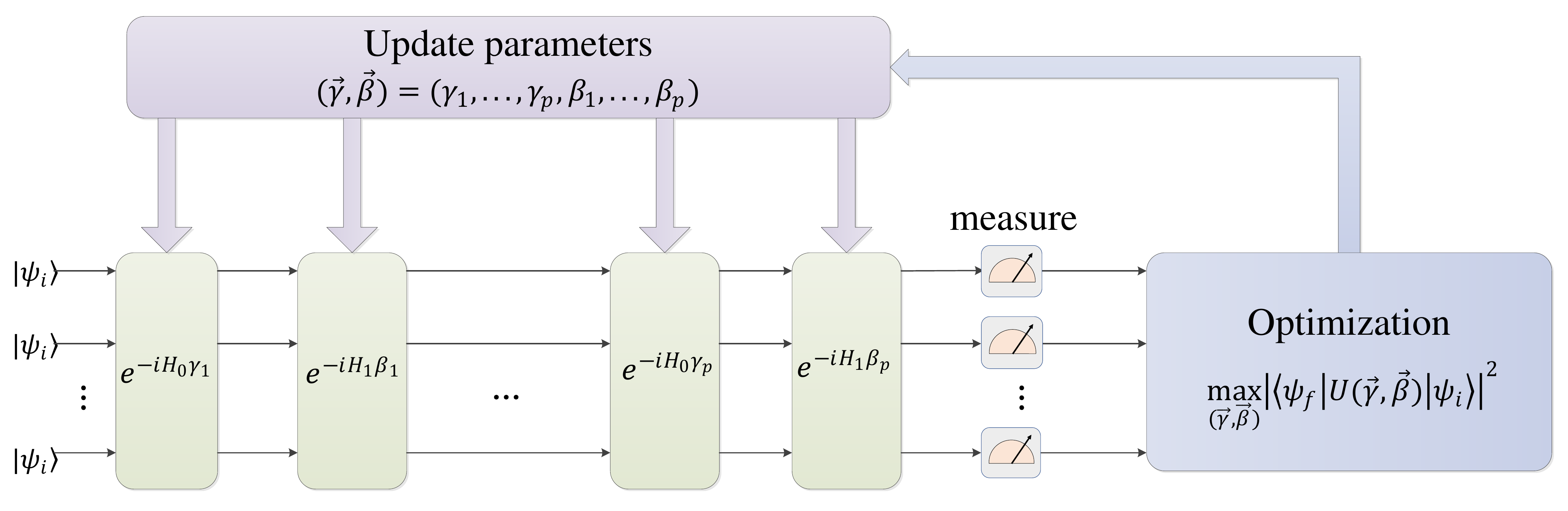}}
\caption{The idea of QAOA. The parameterized unitary controls are implemented on near-term quantum devices to transfer between the states. The fidelity $|\bra{\psi_f} U(\vec{\gamma},\vec{\beta}) \ket{\psi_i}| ^2$ between the initial and target states is obtained by repeating the experiments and measuring the final states. The parameters $(\vec{\gamma},\vec{\beta})$ are optimized using classical algorithms.}
\label{fig1}
\end{figure*}

Quantum control is fundamental for quantum communication and scalable quantum computation\cite{ref.1,ref.2}. State steering of quantum systems \cite{ref.4,bukov2018reinforcement,ref.6,ref.7,ref.8,ref.9} and Quantum Approximate Optimization Algorithm (QAOA) \cite{ref.10,Kusyk21,pan2022automatic} are two major applications of quantum control. These control tasks are often achieved via the application of a series of unitary transformations driven by control Hamiltonians. For example, QAOA switches between two noncommuting Hamiltonians to ensure the universal controllability of quantum states, and the control durations are used to parameterize the quantum control actions (Fig. \ref{fig1}).

However, noise exists in all parts of a quantum system \cite{yao2020policy}, which may have an negative impact on control precision. One of the most common types of noise is the Hamiltonian uncertainty, which means the Hamiltonian may be perturbed as $H(\delta)=\bar{H}+\delta\tilde{H}$, with $\tilde{H}$ being the unknown perturbing Hamiltonian and $\delta$ being the magnitude of perturbation. Previous works have studied the robust control against this kind of uncertainty, such as the gradient-based methods with robust optimization \cite{ref.13,ref.14,ref.31}. Alternatively, Reinforcement Learning (RL) algorithms have also been applied to solve the robust quantum control problem \cite{ref.4,bukov2018reinforcement,ref.6,ref.7,yao2020policy,ref.15,ref.16,ref.17,zhang2019does,ref.24,ref.25,ref.26,ref.27,wauters2020reinforcement,ref.29,Li2020QuantumRL}. During the training process, the agent learns to maximize the reward by measuring the state of the quantum system during, or at the end of the process and taking corresponding control actions.

All the aforementioned algorithms have made use of the full information of the quantum states, which is not available in practice unless a large number of quantum measurements are made. Most of the existing works have obtained the states for optimization by classically simulating the system dynamics, which is clearly not scalable. In essence, the RL control model for noisy quantum systems is expected to be trained and implemented without knowing the underlying dynamical model of the system and noise, which makes it scalable and adaptive to noise. To do this, RL has to rely on measurements of the state which requires huge amount of resources. Therefore, a simplified measurement scheme will hasten the application of RL control on near-term quantum devices. Although a few works \cite{ref.11,wauters2020reinforcement} have considered using partial observations to reduce the resource requirement for the implementation of RL in quantum control tasks, the rewards in these models are still defined in terms of the state fidelity or the expectation of a target Hamiltonian, which still requires an enormous number of quantum measurements to determine the intermediate states. Besides, partial observability can also stem from noisy measurements of the quantum states, which is similar to noisy sensor readings in the classical case. For example, the RL-enhanced QAOA discussed in \cite{ref.11} made use of noisy measurements on the full set of observables and achieved approximation ratios between 0.8 and 0.93, which are taken as acceptable solutions.

Inspired by classical RL algorithms based on Partially Observable Markov Decision Process (POMDP) \cite{ref.37,ref.38,ref.39,ref.41,ref.43}, we propose an RL scheme that uses only partial observations for control decision making and in the meantime avoids using any reward function that employs additional quantum measurements. During the control process, the agent receives partial observations about the current state of the quantum system, and then an optimal action is determined solely based on these observations. As a result, there is no need to estimate the reward with additional quantum measurements, which significantly reduces the measurement and computational cost. We demonstrate this algorithm on two illustrating quantum robust control problems, by imposing noise perturbation to the control Hamiltonian.

This paper is organized as follows. Section \uppercase\expandafter{\romannumeral2} provides a brief introduction to the control model and the definition of partial observations. Section \uppercase\expandafter{\romannumeral3} introduces the quantum RL algorithm with partial observations. Section \uppercase\expandafter{\romannumeral4} presents the numerical results on single-qubit and multi-qubit systems. Section \uppercase\expandafter{\romannumeral5} concludes the paper.

\section{Preliminaries}

\subsection{Control Model}

Consider the Hilbert space $\mathcal{H}=\mathbb{C}^{2^N}$. The $N$-qubit pure quantum state is defined as a complex-valued unit vector in $\mathcal{H}$, where $N$ is the number of qubits. Starting from an initial state $\ket{\psi_i}\in\mathcal{H}$, we apply the following control sequence
\begin{equation}
\begin{split}
\ket{\psi}&=U(\vec{\gamma},\vec{\beta})\ket{\psi_i} \\
&=e^{-iH_1\beta_p}e^{-iH_0\gamma_p}...e^{-iH_1\beta_1}e^{-iH_0\gamma_1}\ket{\psi_i},\label{eq:1}
\end{split}
\end{equation}
where $H_0$ and $H_1$ are noncommuting control Hamiltonians and $p$ is the control depth. For quantum state transfer problem, the objective is to choose the correct control actions $(\vec{\gamma},\vec{\beta})$ to steer the state from $\ket{\psi_i}$ to a target $\ket{\psi_f}$. The existing quantum RL control schemes use the following state fidelity
\begin{equation}
|\langle\psi_t|\psi_f\rangle|^2
\label{eq:2}
\end{equation}
to define the reward at each intermediate step $t$, which requires a huge amount of quantum measurements or accurate simulation of the system dynamics. In contrast, the intermediate states will only be partially observed for the implementation of RL in this paper.

For QAOA, the control scheme is similar as shown in Fig. \ref{fig1}, except that the objective function may be slightly different. 

\subsection{Bloch Sphere Representation of Quantum States}

Since the pure quantum state is represented by a complex-valued vector $\ket{\psi}$ in $\mathcal{H}$, the space of quantum states can be visualized as a generalized unit sphere called Bloch sphere. Consequently, the coordinates for each state can be calculated by projecting the state vector onto an orthogonal basis. Denote the single-qubit Pauli operators as
\begin{equation}
\sigma_x
=
\begin{pmatrix}
 0& 1 \\
 1& 0
\end{pmatrix},\;
\sigma_y
=
\begin{pmatrix}
 0& -i \\
 i& 0
\end{pmatrix},\;
\sigma_z
=
\begin{pmatrix}
 1& 0 \\
 0& -1
\end{pmatrix}, \label{eq:3}
\end{equation}
and the basis states of a single qubit as
\begin{equation}
\ket{0}=(1\quad 0)^T,\ket{1}=(0\quad 1)^T.\label{eq:4}
\end{equation}

\begin{figure}[H]
\centering
\includegraphics[width=8.5cm]{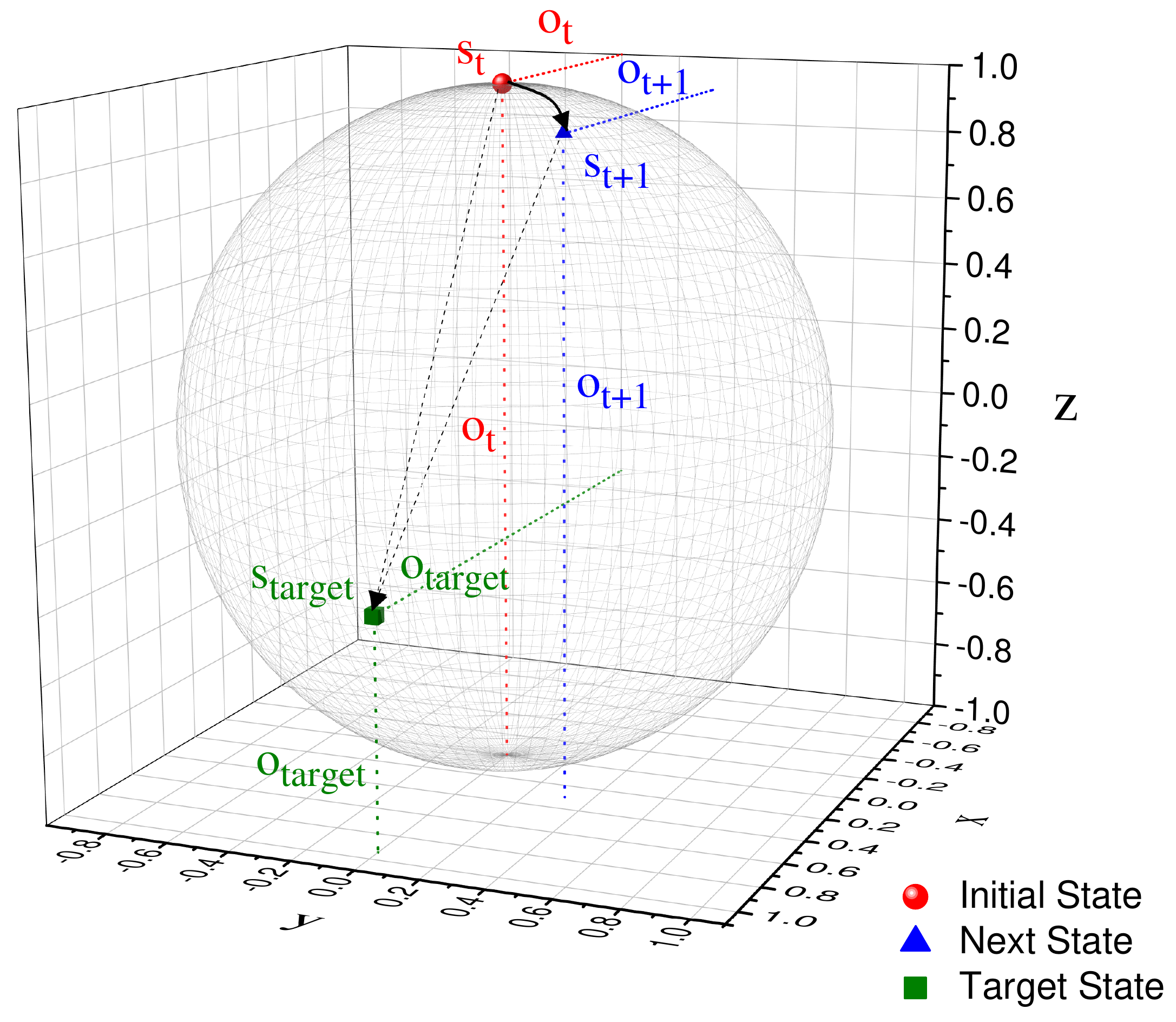}

\caption{Representation of quantum state on the Bloch Sphere. The quantum state moves to the next position under the action of unitary control. Part of the coordinates of the state are observed to calculate the reduced distance between the current state and the target state.}
    \label{fig2}
\end{figure}

It is clear that the Pauli operators span the space of one-qubit states, and thus the coordinates of a state vector $\vec{x}$ on the Bloch sphere can be derived by
\begin{equation}
x_1=\bra{\psi} \sigma_x \ket{\psi},\;
x_2=\bra{\psi} \sigma_y \ket{\psi},\;
x_3=\bra{\psi} \sigma_z \ket{\psi},\label{eq:5}
\end{equation}
with $|x_1|^2+|x_2|^2+|x_3|^3=1$. It should be noted that Eq.~(\ref{eq:5}) is the mathematical formulation of quantum measurements, with the Pauli operators being the measurement operations and $\ket{\psi}$ being the state to be measured. $x_1,x_2,x_3$ are the corresponding measurement results which are obtained by averaging the measured outcomes on a large number of copies of $\ket{\psi}$. The quantum state is fully observable if all three observations in Eq.~(\ref{eq:5}) can be made.

The coordinates of multi-qubit states can be defined in a similar way. Denote $\sigma_0 = I$. The coordinates of a multi-qubit state are given by
\begin{equation}
\bra{\psi} \underbrace{\sigma_{k_1} \otimes \sigma_{k_2} \otimes ... \otimes \sigma_{k_N}}_{N} \ket{\psi},\quad k_n=0,x,y,z \label{eq:6}
\end{equation}
except for $k_1=k_2=...=k_N=0$. Hence, the dimension of an $N$-qubit pure state vector is $4^N-1$. In this paper, partial observations are made on a subset of these generalized Pauli operators.

\begin{figure*}[htbp]
\centering
\includegraphics[height=9cm]{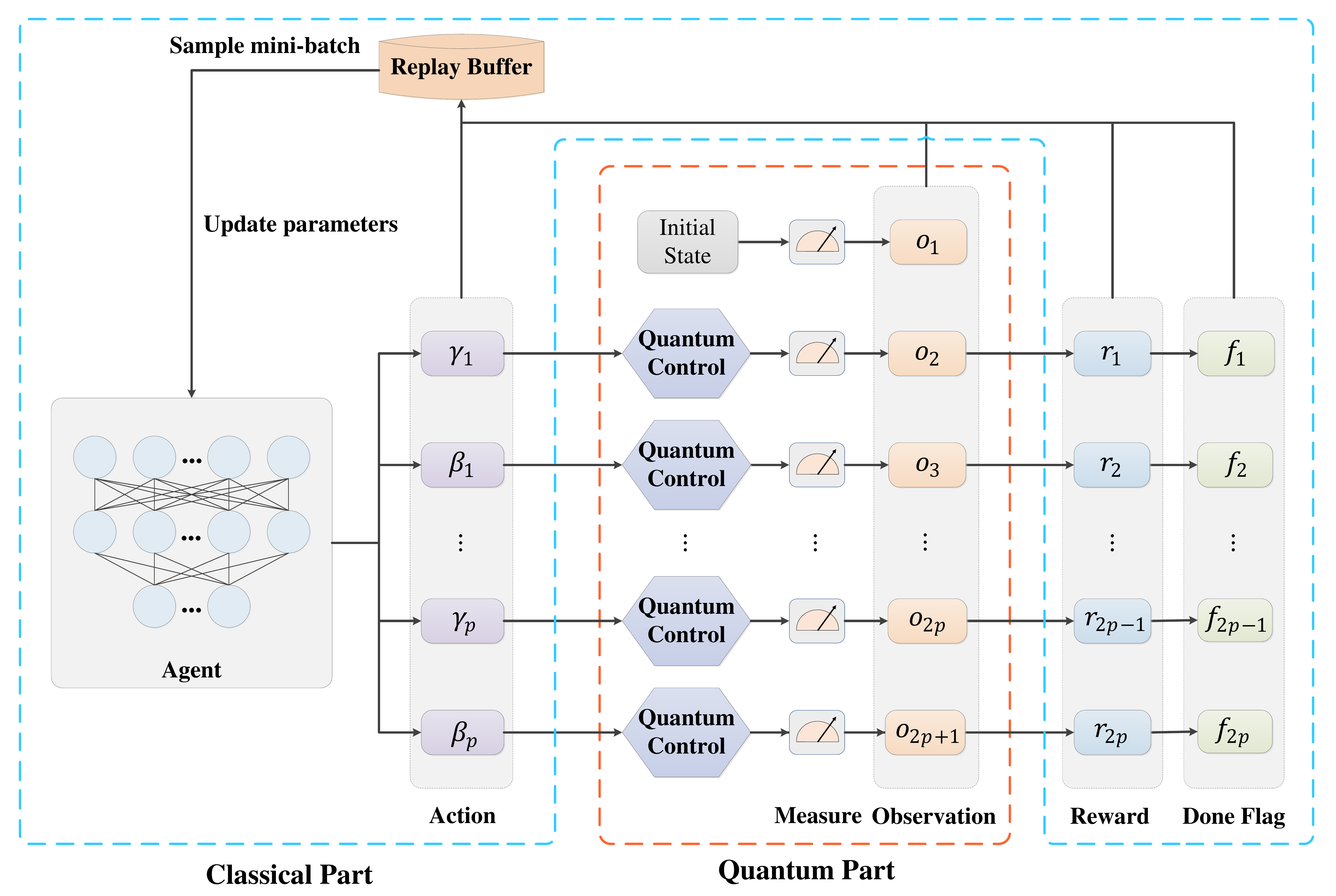}

\caption{The diagrammatic representation of the quantum RL control algorithm with replay buffer. The agent generates an action which is either $\gamma_j$ or $\beta_j$, depending on the alternating control Hamiltonian chosen for the current step. This process is repeated until $t=2p$. Experience tuples $(\vec{o}_t,a_t,r_t,\vec{o}_{t+1},f_t)$ are stored into the replay buffer. A subset of tuples is sampled from the buffer for updating the agent.}
    \label{fig3}
\end{figure*}

\begin{figure*}[htbp]
\centerline{\includegraphics[height=9cm]{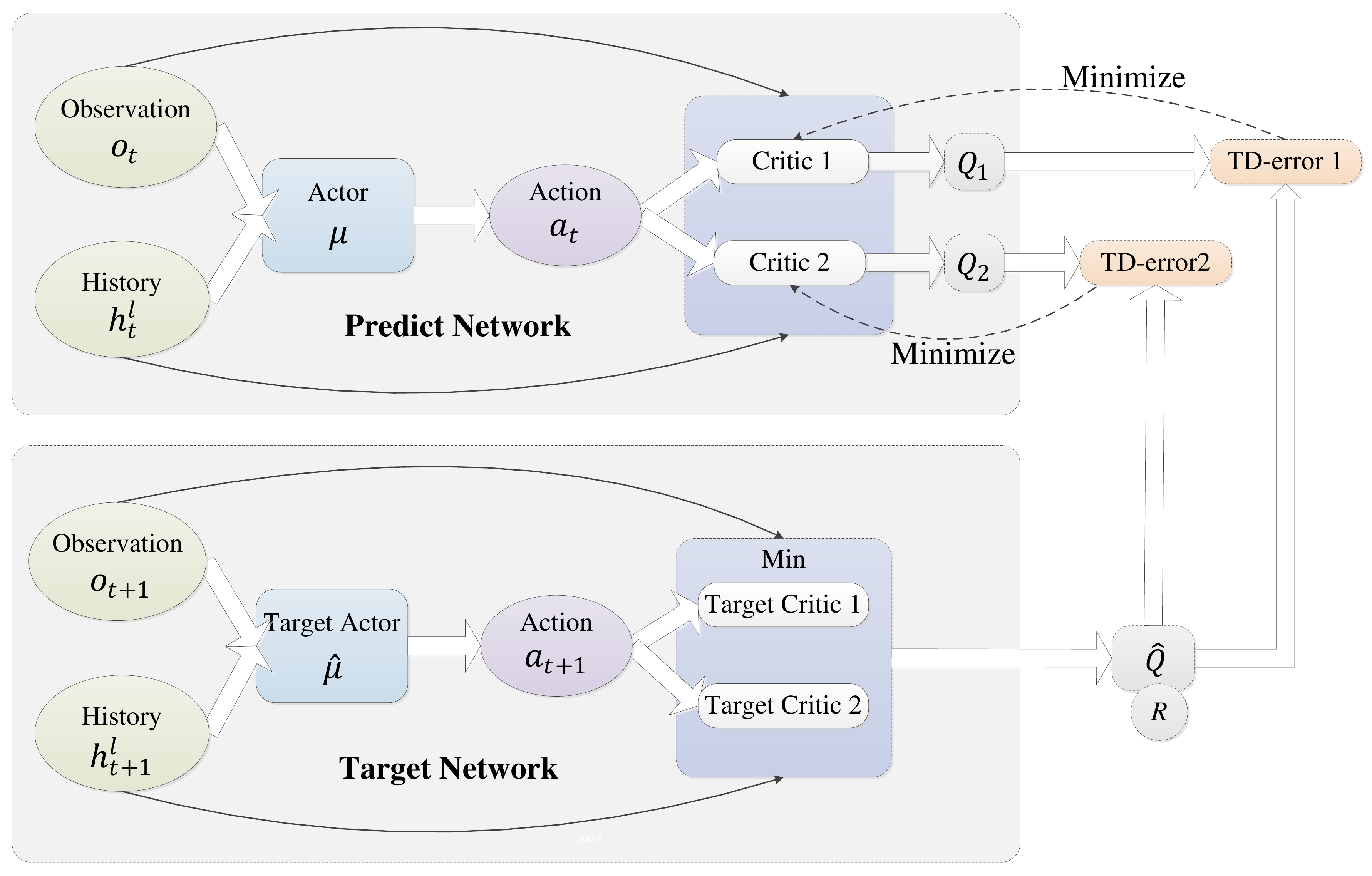}}
\caption{The framework of TD3. At the start of the training, the actor and two critics of the predict network are randomly initialized. Then their parameters are copied to the target networks. In target network, the target Q-value $\hat{Q}$ is chosen as the smaller one of the estimated optimal Q-values of two target critics to avoid overestimation. The history $h_t^l$ contains $l$ pairs of past observations and actions as $\{\vec{o}_i,a_i\},i=t-l,...,t-1$. $\mu(\vec{o}_t,h_t^l)$ is an actor which generates the action $a_t$ based on $\vec{o}_t$ and the history $h_t^l$. $Q(\vec{o}_t,a_t,h_t^l)$ denotes the estimated Q-value. }
\label{fig5}
\end{figure*}

\section{Quantum RL Algorithm with Partial Observation}

\subsection{Model Description}
First, we introduce the quantum RL control scheme of this paper.

\begin{itemize}
\item \textbf{State}. The quantum state is defined by the coordinate vector $\vec{x}$ on the Bloch sphere. Due to the noise and uncertain environments, the agents may only receive partial observations of the current state.

To be more specific, according to the control model (\ref{eq:1}), the quantum state evolves as
\begin{equation}
\ket{\psi_t}=\left\{
\begin{array}{ll}
e^{-i \gamma_t (H_0+\delta H_0)} \ket{\psi_{t-1}} &\mbox{if}\ t\ \mbox{is\ odd},\\
e^{-i \beta_t (H_1+\delta H_1)} \ket{\psi_{t-1}} &\mbox{if}\ t\ \mbox{is\ even}.
\end{array} \right.\label{eq:7}
\end{equation}
Due to the unknown perturbations $\delta H_0$ and $\delta H_1$, the quantum state at each time step $t$ is hidden to the agent unless a full observation is conducted.

\item \textbf{Partial Observation}. The quantum observation is made by applying quantum measurements to the states with respect to a subset of generalized Pauli operators, which will give the estimates $\vec{o}_t$ for only a part of the coordinates of the state. Alternatively, the full set of coordinates may be observed, with a fraction of the observations $\vec{o}_t$ being single-shot or extremely noisy.

\item \textbf{Reward}. As depicted in Fig.~\ref{fig2}, the reward for optimizing the quantum control action is defined as the reduced distance between the partially observed coordinates of the current state and target state on the Bloch sphere, which is calculate by $r_t = \sigma(|\vec{o}_t- \vec{o}_{target}|-|\vec{o}_{t+1}-\vec{o}_{target}|),\sigma>0$. 

\item \textbf{Action}. The action $a_t$ is chosen from a continuous interval, representing the values of the control durations (or angles) $\gamma_j,\beta_j$. Similar to the classical RL, the actor generates the action $a_t$ through an actor-critic framework given the observations as input. After that, the action is used to parameterize the next unitary control.

\end{itemize}

The general framework for the quantum RL control is depicted in Fig.~\ref{fig3}. At the time step $t$, the agent receives the partial observation $\vec{o}_{t+1}$ which allows us to calculate the reward $r_t$. Here we use a done flag to mark the end of each trajectory. If $t=2p$, the done flag $f_t$ is set to 1, otherwise it is set to 0. After taking the action $a_t$, the experience $(\vec{o}_t, a_t, r_t, \vec{o}_{t+1}, f_t)$ is sent into the replay buffer which records the historical data. The parameters of the agent are optimized by sampling mini-batches from the replay buffer. More details on POMDP can be found in Appendix.~\ref{secpomdp}.

\subsection{The Implementation of Robust Control using Memory-based TD3 framework}
Any established RL-based frameworks can be used to train the agent as defined in Fig.~\ref{fig3}. Here we adopt the TD3 deep neural networks from \cite{ref.51}. TD3 is based on the commonly-used actor-critic structure as depicted in Fig.~\ref{fig5}, which involves a total of two actor networks and four critic networks. TD3 is said to be able to solve the problem of overestimating the Q-value. During the training process, each critic in the predict network is optimized to minimize the mean-square-error between the predicted $Q_{j\in\{1,2\}}$ and the estimated target $\hat{Q}$. The pseudo-code that implements the quantum robust control using TD3 is shown in Algorithm \ref{alg:1}.

\renewcommand{\algorithmicrequire}{\textbf{Input:}}
\begin{breakablealgorithm}
    \caption{Pseudo-code for Quantum Robust Control with Partial Observations}
    \begin{algorithmic}[1]
    	\Require History length $L$
        \State Initialize critics $Q_{\theta^{Q_1}}$, $Q_{\theta^{Q_2}}$, and actor $\mu_{\theta^{\mu}}$ with random parameters $\theta^{Q_1}$, $\theta^{Q_2}$ and $\theta^{\mu}$
        \State Initialize target networks $\theta^{Q^-_1}\gets\theta^{Q_1}$, $\theta^{Q^-_2}\gets\theta^{Q_2}$ and $\theta^{\mu^-}\gets\theta^{\mu}$
     	\State Set the initial state and initialize environment $\vec{o}_1=$ env.reset(), past history $h^l_1\gets0$, perturbation strength $\delta$ and replay buffer $D$
        \For {$t=1$ to $T$}
            \State Select action with exploration noise
            \par
            $a_t\sim\mu_{\theta^{\mu}}(\vec{o}_t,h^l_t)+\epsilon$, $\epsilon\sim\mathcal{N}(0,0.1)$
            \State Compute the next state $\ket{\psi_{t+1}}$ according to $\ket{\psi_{t+1}}=\exp\{-iH(\delta) a_t\}\ket{\psi_{t+1}}$, with $H(\delta)=\bar{H}+\delta\tilde{H}$
            \State Make the partial observation $\vec{o}_{t+1}$ on $\ket{\psi_{t+1}}$
            \State Compute the reward $r_t=\sigma(|\vec{o}_t - \vec{o}_{target}| - |\vec{o}_{t+1}-\vec{o}_{target}|)$
            \If{$t==2p$}
            	\State This trajectory ends: $f_t=1$
            \Else
            	\State $f_t=0$
            \EndIf
            \State Store the experience tuple $(\vec{o}_t,a_t,r_t,\vec{o}_{t+1},f_t)$ into $D$
            \If{$f_t$}
        		\State Reset environment $\vec{o}_{t+1}=$ env.reset() and history $h^l_{t+1}\gets0$
    		\Else		
        		\State $h^l_{t+1}=(h^l_t-(\vec{o}_{t-l},a_{t-l}))\cup(\vec{o}_t,a_t)$
    		\EndIf
        	\State Sample a mini-batch of experiences with histories as $\{(h^l_t,\vec{o}_t,a_t,r_t,\vec{o}_{t+1},d_t)_i\}$ from $D$
        	\State Optimize $Q_j$ using the mini-batch
            \State Optimize $\mu$ using the mini-batch
            \State Update the target actor and critics by copying from the predict network
		\EndFor
    \end{algorithmic}
    \label{alg:1}
\end{breakablealgorithm}

Since any RL-based frameworks can be readily used to implement the proposed algorithm of this paper as an alternative, we do not intend to expand on the design of the deep neural network structures of TD3. Further details on the exact implementation of the neural networks can be found in Appendix.~\ref{secrl} and \cite{ref.51}.

In particular, step 6 in Algorithm \ref{alg:1} will be replaced by real quantum control in practical implementation on quantum hardware. In that case, Algorithm \ref{alg:1}, which is solely based on partial measurements, can still work with unknown perturbations and noises. However, if prior knowledge on the source of the perturbation and noise is available, the RL control model can be pretrained with classical simulations, which will provide significant cost reduction since experimenting on quantum hardware is very expensive at present.

\section{Numerical Experiments}
In this section, the performances of the partially observed RL algorithm are first demonstrated on the robust state control problem of single- and multi-qubit systems. Then we also test the algorithm on multi-qubit QAOA.

\subsection{Single-qubit Case}

For single-qubit robust control, the number of control steps is chosen to be 20. We adopt the experimental setup from \cite{ref.15}, in which the initial and target states are given by
\begin{equation}
\ket{\psi_i}=\ket{0},\quad \ket{\psi_f}=\frac{1}{\sqrt{3}}\ket{0}+\frac{\sqrt{2}}{\sqrt{3}}\ket{1},\label{eq:8}
\end{equation}
and the control Hamiltonians are given by
\begin{equation}
H_0= -\frac{1}{2}\sigma_z+2\sigma_x,\quad H_1= -\frac{1}{2}\sigma_z-2\sigma_x.  \label{eq:9}
\end{equation}
The perturbations $\delta \tilde{H}_0$ and $\delta \tilde{H}_1$ to $H_0$ and $H_1$ are randomly generated for each trajectory of training. That is, at the beginning of each trajectory of training, we randomly generate the diagonal elements of $\delta H_0$ and $\delta H_1$ from a uniform distribution $U(0,1)$. The real and imaginary parts of the elements of the upper triangle of $\delta H_0$ and $\delta H_1$ are sampled from $U(0,1)$ as well. The lower triangle is then generated as the complex conjugate of the upper triangle to guarantee that the perturbed Hamiltonian is Hermitian. $\delta$ is a positive parameter that controls the magnitude of perturbation. The action values are confined within $\mathcal{A}=[-5,5]$, and the initial values of actions for each trajectory is uniformly sampled from $\mathcal{A}=[-5,5]$. The mini-batch size is set as $100$ and the maximal length of history is set as $20$ for the simulations. The parameters of the agent are optimized using the Adam optimizer with a learning rate of $0.00001$ for actor and critics. Also, we set $\sigma=10$ in the reward function. We assume that the partial observations are made on the observables $\sigma_x$ and $\sigma_z$. Therefore, the observations obtained at time step $t$ are calculated by $\mathcal{O} = \{\bra{\psi_t} \sigma_x \ket{\psi_t}, \; \{\bra{\psi_t} \sigma_z \ket{\psi_t}\}$.

\begin{figure}[H]
\centering
\subfigure[]{
\includegraphics[width=9cm]{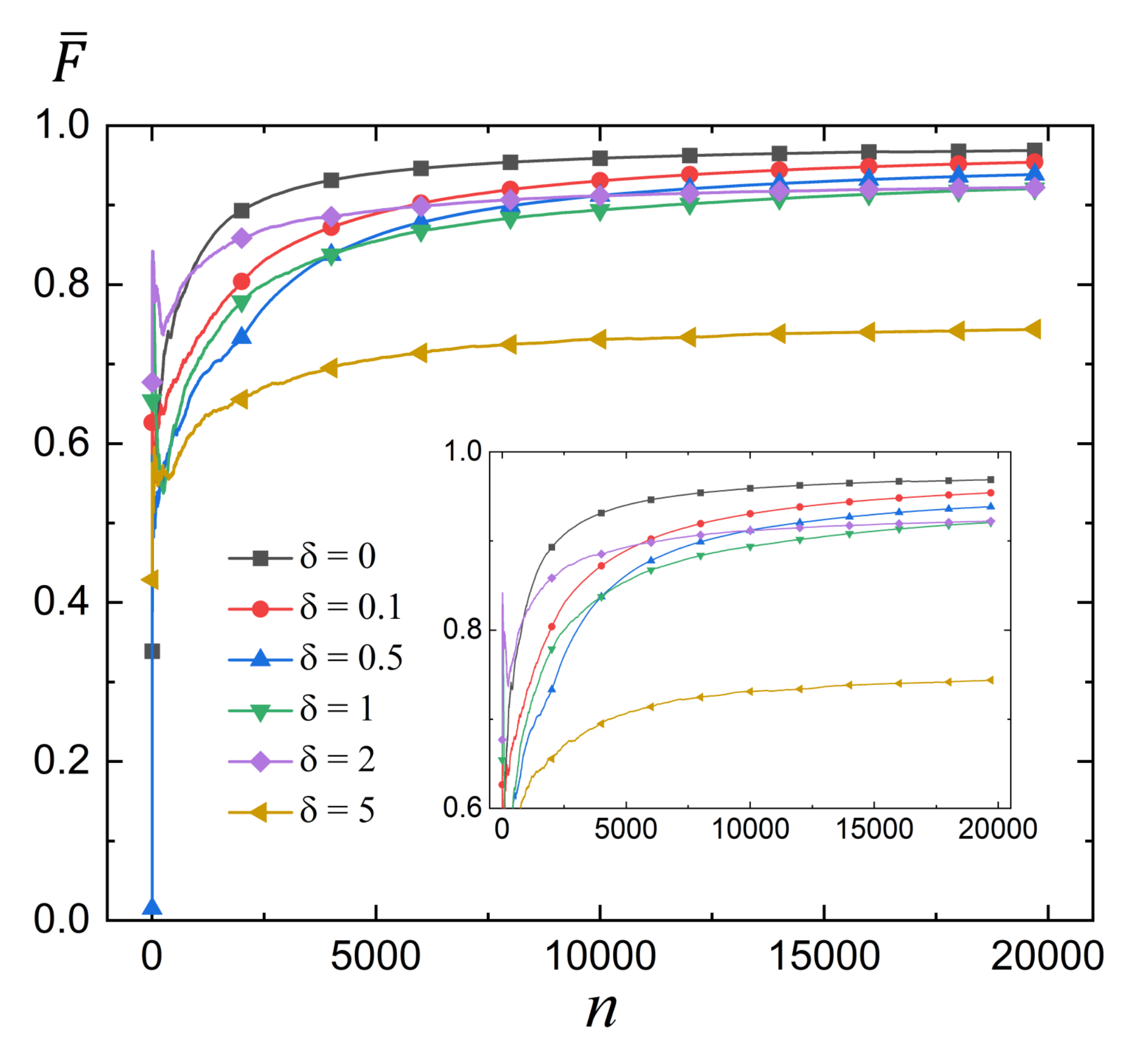}
}
\subfigure[]{
\includegraphics[width=9cm]{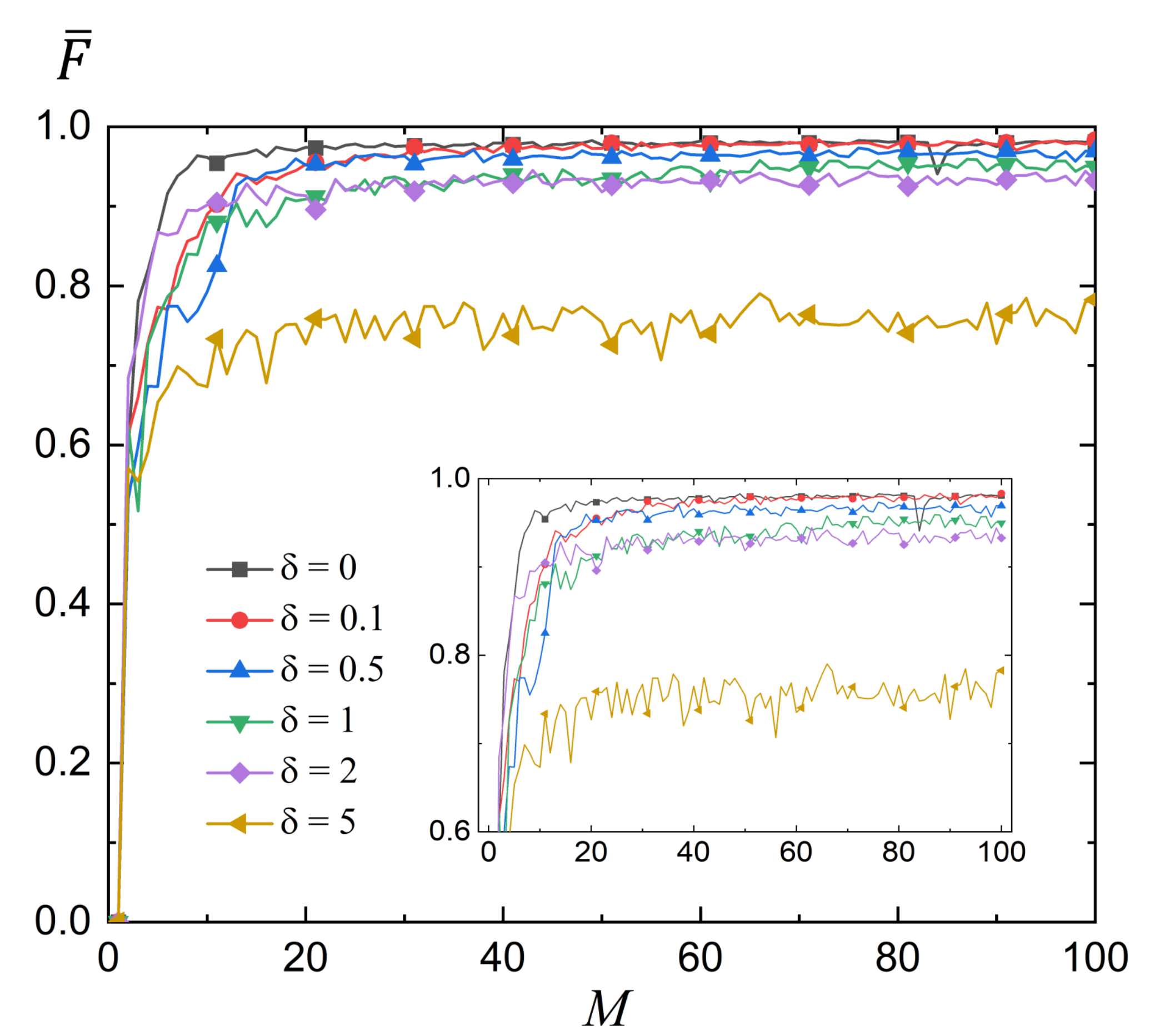}
}
\caption{The evolution of average fidelity during the training. Curves with different colors correspond to different level of perturbations. (a) shows the average fidelities achieved by the current $n$ trajectories. (b) shows the average fidelities within each epoch, where each epoch contains 200 trajectories of training. As a result, 20000 trajectories of training in (a) can be divided into a total of $100$ epochs in (b).}
    \label{fig6}
\end{figure}

The average fidelities between the initial and target states during the training are shown in Fig.~\ref{fig6}(a). The model is trained for 20000 times, corresponding to 20000 complete trajectories. As $\delta$ increases, the average 

\begin{figure}[H]
\centering
\subfigure[]{
\includegraphics[width=8.5cm]{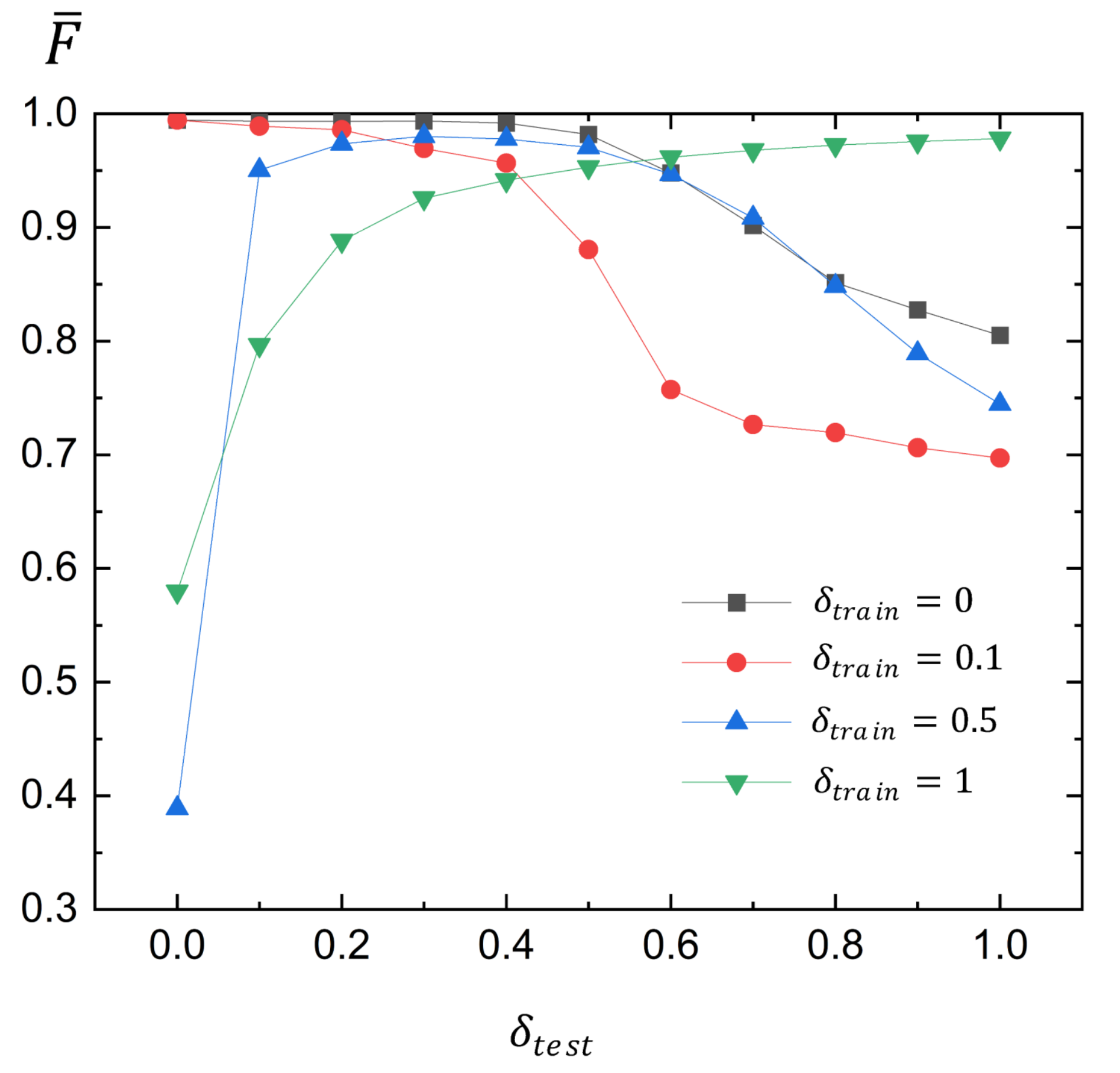}
}
\subfigure[]{
\includegraphics[width=8.5cm]{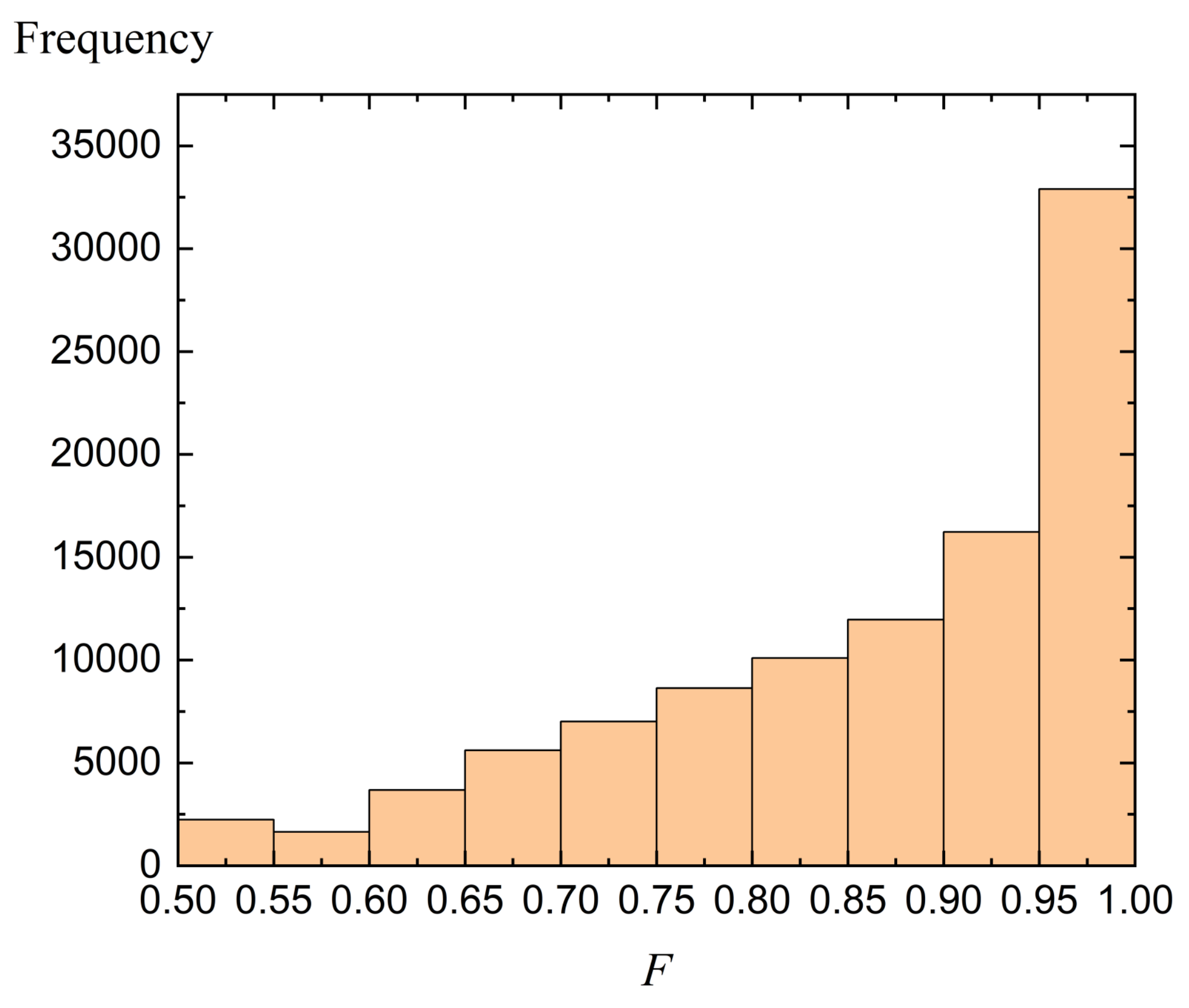}
}
\caption{Numerical results on the generalizability of our method. (a) The trained models are tested with different noise levels. $\delta_{train}$ is the noise level for training the model, and $\delta_{test}$ is used for testing the model with the same initial and target states. For each $\delta_{test}$, $10000$ trajectories are generated with random and independent perturbations. Then the average fidelity achieved by these trajectories is calculated. (b) The model trained with $\delta=0.1$ is tested on random initial and target states. $100000$ pairs of randomly generated initial and target states are tested. The frequency distribution of the achieved fidelities is depicted in this figure.}
\label{fig7}
\end{figure}

\begin{table*}[ht]
\centering
\caption{Comparison with the commonly-used algorithms with full observability for single-qubit case.}
\label{tab1}
\setlength{\tabcolsep}{8pt}
\begin{tabular}{|c|c|c|}
\hline
\makecell[c]{Algorithms}    &    \makecell[c]{Average Fidelity}    &    \makecell[c]{Standard Deviation}\\
\hline
\makecell[c]{SGD ($\delta=0$)}     &    \makecell[c]{0.950333}    &    \makecell[c]{0.072376}\\
\hline
\makecell[c]{Krotov ($\delta=0$)}   &    \makecell[c]{0.989746}    &    \makecell[c]{0.021458}   \\
\hline
\makecell[c]{Deep Q Learning ($\delta=0$)}   &    \makecell[c]{0.991249}    &    \makecell[c]{0.081046}\\
\hline
\makecell[c]{Policy Gradient ($\delta=0$)}    &    \makecell[c]{\bf{0.999562}}    &    \makecell[c]{0.002871}\\
\hline
\makecell[c]{Our Method ($\delta=0$)}    &    \makecell[c]{0.998972}    &    \makecell[c]{\bf{0.001112}}\\
\hline
\makecell[c]{Our Method ($\delta=0.1$)}    &    \makecell[c]{0.989171}    &    \makecell[c]{0.014606}\\
\hline
\makecell[c]{Our Method ($\delta=0.5$)}    &    \makecell[c]{0.979764}    &    \makecell[c]{0.039399}\\
\hline
\makecell[c]{Our Method ($\delta=1$)}    &    \makecell[c]{0.973757}    &    \makecell[c]{0.088222}\\
\hline
\end{tabular}
\end{table*}

\noindent
fidelities show an obvious decrease due to the random uncertainties in the control Hamiltonian. Particularly, the fidelities averaged over a local epoch of trajectories are more useful to demonstrate the final performance of the algorithm. In Fig.~\ref{fig6}(b), the fidelities are calculated after every 200 trajectories, and we can see that the average fidelity can reach above 0.98 when $\delta=0$ or $\delta=0.1$. The ultimate fidelity with $\delta=0.5$ is still higher than 0.96. If we increase $\delta$ to $2$, which makes the amplitude of perturbation approximately equal to the amplitude of control Hamiltonians, the algorithm can achieve an average fidelity between 0.93 and 0.94. It should be noted that this noise level is already much larger than the actual noise level in quantum computing systems.

Next, we perform additional tests to demonstrate the generalizability of our method. In Fig.~\ref{fig7}(a), the agents are trained with $\delta_{train}=0,0.1,0.5,1$, respectively. Then the parameters of the agents are fixed, and the performances of the models are tested with different $\delta$ on the same initial and target states. An interesting observation is that models trained with higher level of noise ($\delta_{train}=0.5,1$) show better generalizability than models trained with little or no noise ($\delta_{train}=0,0.1$). For example, when $\delta_{train}=1$, the test fidelity can still reach above 0.9 with any $\delta_{test}$ that is larger than 0.3. In contrast, when $\delta_{train}=0.1$, the test fidelity quickly drops below 0.9 before $\delta_{test}=0.5$. This observation suggests that adding a certain amount of noise during training is beneficial for improving the robustness and generalizability of the model.

We have also tested the performance of the trained model on different initial and target states. In Fig.~\ref{fig7}(b), the agent is trained with $\delta=0.1$, and $100000$ pairs of initial and target states are randomly generated for test by
\begin{equation}
\ket{\psi}=\alpha \ket{0} + \beta \ket{1},\label{eq:10}
\end{equation}
with
\begin{equation}
\alpha \sim U(0,1),\quad \beta = \sqrt{1 - \alpha^2}.\label{eq:11}
\end{equation}
The test experiments are still conducted with $\delta=0.1$. It is clearly that the trained model works well with unseen initial and target states, with more than 32500 pairs of states having achieved a final fidelity above 0.95.

Our method is then compared to other commonly-used algorithms in Table~\ref{tab1} under the same experimental setup with full observability. These algorithms consist of two gradient-based methods: stochastic gradient descent (SGD) \cite{ferrie2014self}, Krotov algorithm \cite{morzhin2019krotov} and two commonly-used RL algorithms: deep Q-learning \cite{bukov2018reinforcement} and deep policy gradient \cite{yao2020policy,zhang2019does}. SGD is one of the simplest gradient-based optimization algorithms, in which the control is updated using the approximated gradient of the cost function based on random batches. In Krotov algorithm, the initial state is firstly propagated forward to obtain the evolved state. Then the evolved state is projected to the target state, defining a co-state that encapsulates the mismatch between the two. Finally, the co-state is propagated backward to the initial state, during which the controls are updated. As this process converges, the co-state becomes identical to the target state. For comparison, each algorithm has been run 100 times to calculate the average fidelity without assuming any noise or Hamiltonian uncertainty, with codes adopted from \cite{zhang2019does}. The achieved fidelity of our method ($\delta=0$) is higher than those of other algorithms including deep Q-Learning when noise is not taken into consideration. If noise is added to the control Hamiltonian of our method, e.g. $\delta=0.1$ or $\delta=0.5$, the performance of our method is still better than the conventional SGD and Krotov methods and similar to deep Q-learning and policy gradient method, which did not consider any noise effect. These numerical results demonstrate that our method can at least achieve the same level of performance in comparison with the state-of-the-art algorithms using only partial observations while being adaptive to noise.

It should be noted that we have $|\vec{x}|^2=1$, or that the control landscape is a unit sphere. As a result, the estimates of $\langle \sigma_x\rangle$ and $\langle \sigma_z \rangle$ can be used to jointly determine the value of $|\langle \sigma_y\rangle|$ which is unobserved. The RL model may have learned to infer this hidden relation with partial observations to promote the control precision.

\subsection{Two-qubit Case}

Consider the commonly-used spin-chain example for multi-qubit control \cite{wauters2020reinforcement}, in which the control Hamiltonians are defined as
\begin{equation}
H_0= -J \sum_{j=1}^{N-1} \sigma_j^z \sigma_{j+1}^z \label{eq:12}
\end{equation}
and
\begin{equation}
H_1= -\sum_{j=1}^{N} \sigma_j^x,\label{eq:13}
\end{equation}
where $J$ is the coupling strength. Then initial state is given by
\begin{equation}
\ket{\psi_i}=\frac{1}{\sqrt{2^N}} \otimes_j (\ket{0}_j+\ket{1}_j).\label{eq:14}
\end{equation}
Without loss of generality, we let $N=2$ and $J=0.5$ for the demonstration. 
In this case, the control Hamiltonians and initial state are written as
\begin{equation}
H_0= -\frac{1}{2} \sigma_1^z \sigma_2^z,\quad H_1= -\sigma_1^x-\sigma_2^x\label{eq:15}
\end{equation}
and
\begin{equation}
\ket{\psi_i}=\frac{1}{2} (\ket{00}+\ket{01}+\ket{10}+\ket{11})\label{eq:16}.
\end{equation}

\begin{table*}[ht]
\centering
\caption{Comparison with the commonly-used algorithms with full observability for two-qubit case.}
\label{tab2}
\setlength{\tabcolsep}{8pt}
\begin{tabular}{|c|c|c|}
\hline
\makecell[c]{Algorithms}    &    \makecell[c]{Average Fidelity}    &    \makecell[c]{Standard Deviation}\\
\hline
\makecell[c]{Krotov ($\delta=0$)}   &    \makecell[c]{0.978592}    &    \makecell[c]{0.068521}\\
\hline
\makecell[c]{Deep Q Learning ($\delta=0$)}   &    \makecell[c]{0.990445}    &    \makecell[c]{0.060376}\\
\hline
\makecell[c]{Policy Gradient ($\delta=0$)}    &    \makecell[c]{\bf{0.998939}}    &    \makecell[c]{\bf{0.003715}}\\
\hline
\makecell[c]{Our Method ($\delta=0$)}    &    \makecell[c]{0.984726}    &    \makecell[c]{0.018958}\\
\hline
\makecell[c]{Our Method ($\delta=0.01$)}    &    \makecell[c]{0.981503}    &    \makecell[c]{0.019187}\\
\hline
\makecell[c]{Our Method ($\delta=0.02$)}    &    \makecell[c]{0.965358}    &    \makecell[c]{0.022574}\\
\hline
\makecell[c]{Our Method ($\delta=0.05$)}    &    \makecell[c]{0.936025}    &    \makecell[c]{0.027871}\\
\hline
\end{tabular}
\end{table*}

Here the target state is chosen as the ground state of
\begin{equation}
H_{targ}= H_0+H_1.\label{eq:17}
\end{equation}
The observations at time step $t$ are made as follows
\begin{equation}
\mathcal{O} = \left\{
\begin{array}{ll}
\bra{\psi_t} \sigma^x_1 \otimes I \ket{\psi_t}, &
\bra{\psi_t} \sigma^z_1  \otimes I \ket{\psi_t}, \\
\bra{\psi_t} I \otimes \sigma^x_2 \ket{\psi_t}, &
\bra{\psi_t} I \otimes \sigma^z_2 \ket{\psi_t}, \\
\bra{\psi_t} \sigma^x_1 \otimes \sigma^x_2 \ket{\psi_t}, &
\bra{\psi_t} \sigma^z_1 \otimes \sigma^z_2 \ket{\psi_t}.
\end{array}
\right\}
\label{eq:18}
\end{equation}

\begin{figure}[H]
\centering
\subfigure[]{
\includegraphics[width=9cm]{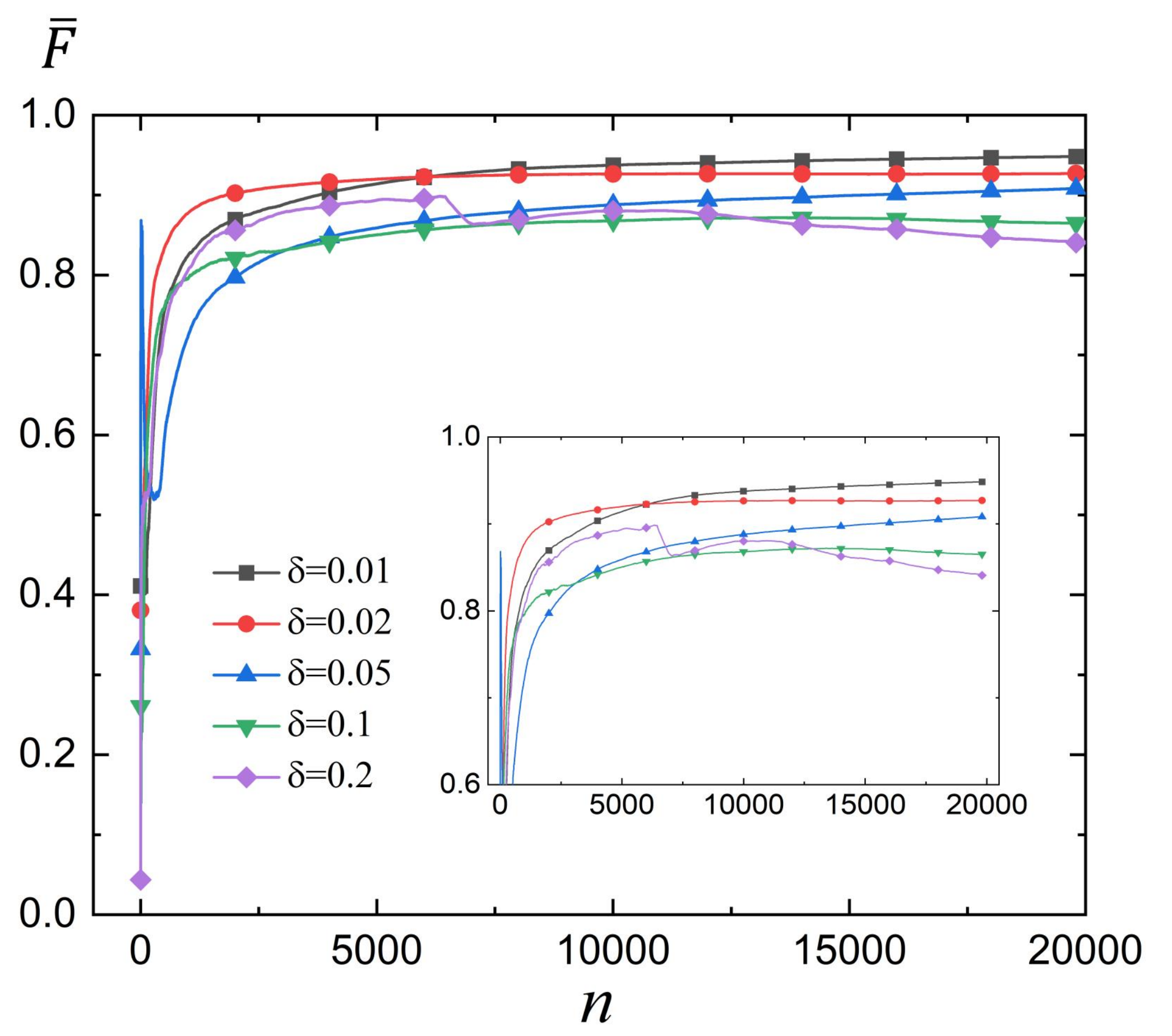}
}
\subfigure[]{
\includegraphics[width=9cm]{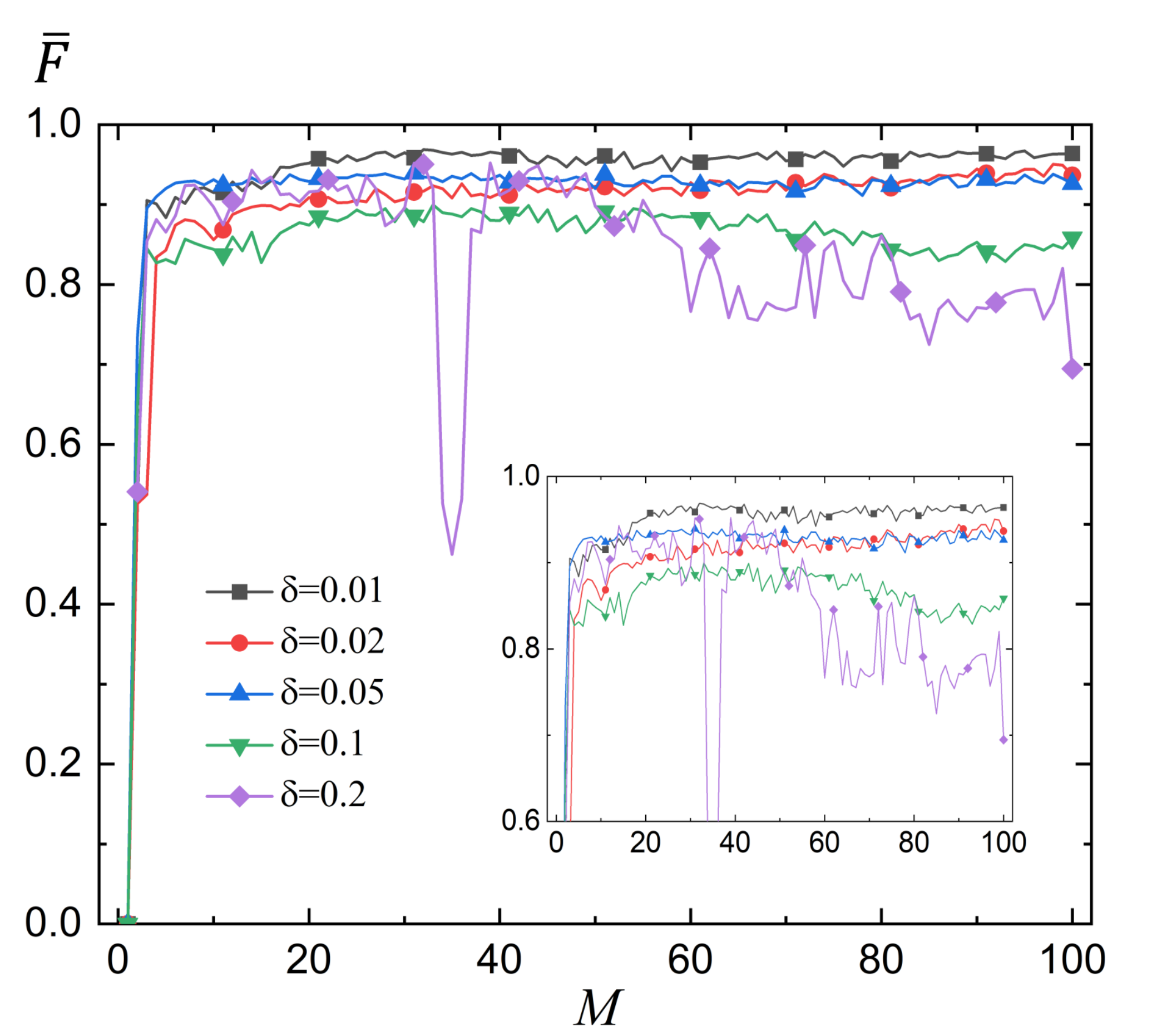}
}
\caption{The average fidelities of the two-qubit state control during the training. (a) shows the training curves of the average fidelities for different levels of perturbations. (b) shows the average fidelity of every 200 trajectories at different noise levels. The total number of epoch is $M=100$.}
\label{fig8}
\end{figure}

The numerical results in Fig.~\ref{fig8} show that the RL algorithm with 6 observations can achieve high-fidelity control against a low level of perturbations to the Hamiltonian. As the amplitude of the perturbations increases, the average fidelities will decrease. In particular, the fluctuations in the ultimate fidelities have become more severe with $\delta=0.2$, which can be seen from Fig.~\ref{fig8}(b). This may be due to the fact that the number of randomly perturbed elements of the control Hamiltonians is growing exponentially with the number of qubits, which indicates an exponential growth in the uncertainties of the Hamiltonians. In addition, the number of partial observations for two qubits is only 6, while the dimension of the state space is $4^2-1=15$. In contrast, 3 observations are enough for the complete determination of a single-qubit state, and thus partial observation with 2 observables will make the single-qubit algorithm more tolerant of a large $\delta$.

\begin{figure}[H]
\centering
\includegraphics[width=9cm]{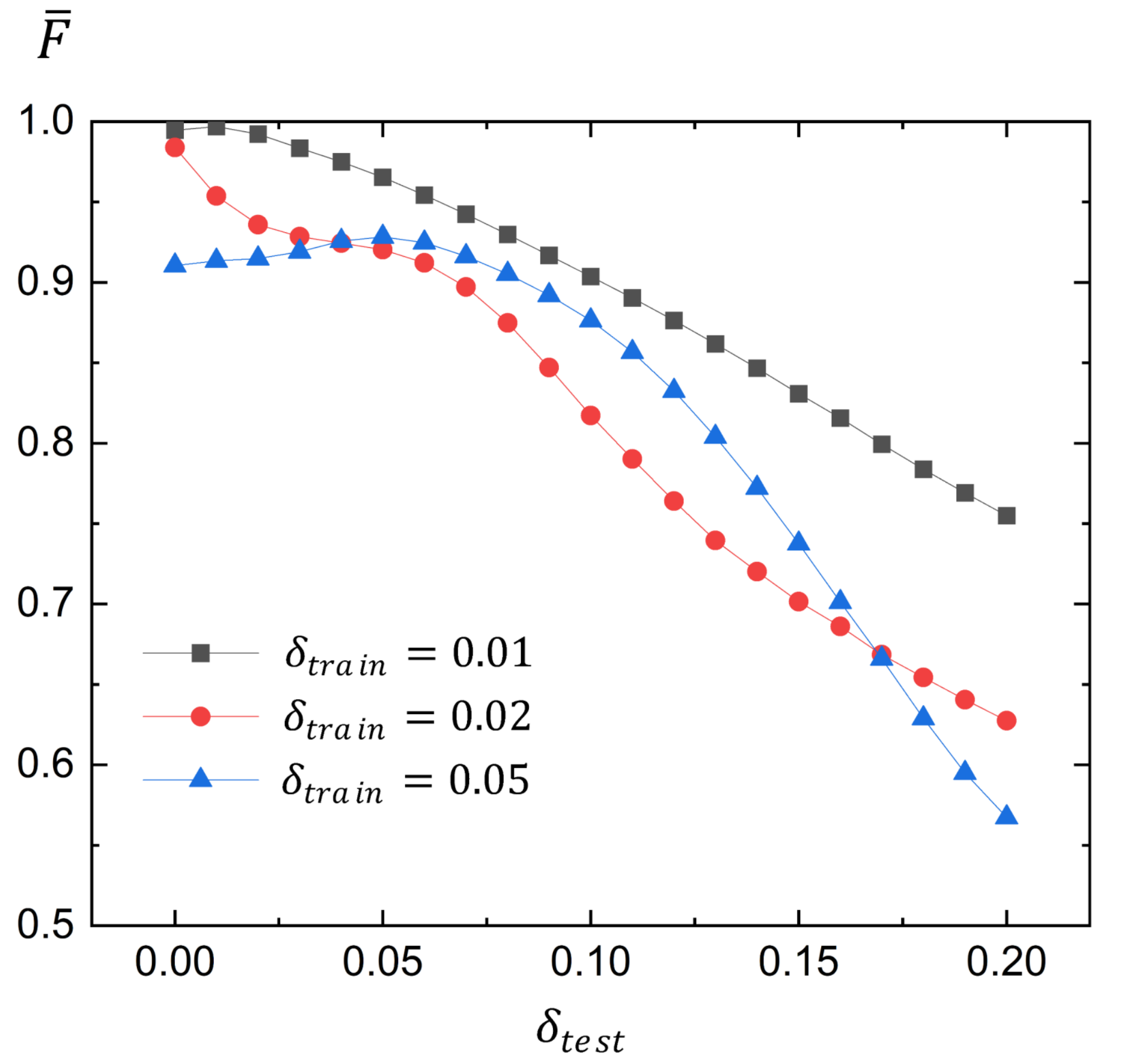}
\caption{The test results of the trained models with different noise levels.}
\label{fig9}
\end{figure}

We also test the trained agents with different noise levels, as shown in Fig.~\ref{fig9}. Since the number of observations is too few to counteract the increased effects of noise in the two-qubit system, the agents suffer from an abrupt performance deterioration if $\delta_{test}$ is much larger than $\delta_{train}$. Similar to the single-qubit example, the trained model with the largest $\delta_{train}=0.05$ has the most stable performance during the test under a small variation of the noise amplitude.

As can be seen from Table~\ref{tab2}, our method does not always outperform the state-of-the-art reinforcement learning algorithms. However, with partial observations, our method can at least achieve the same level of performance in comparison with the state-of-the-art algorithms. SGD is not studied in the two-qubit case since it has been proved to be an inefficient method for solving many-body problems \cite{zhang2019does}.

\begin{figure}[H]
\centering
\includegraphics[width=8.8cm]{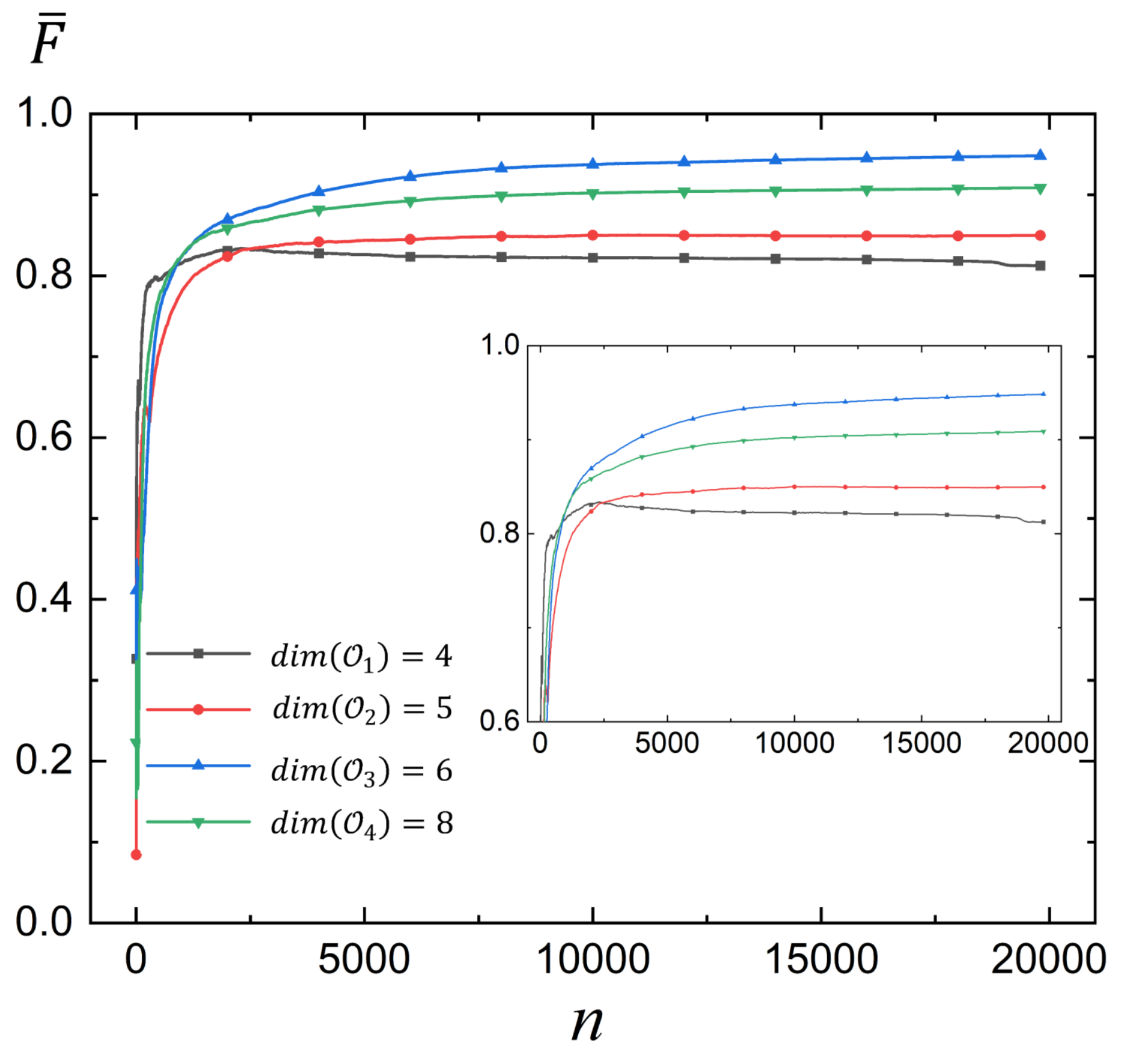}
\caption{The averaged fidelities during training for the two-qubit case with different number of observations.}
\label{fig10}
\end{figure}

Next, we train the agents with 4 different configurations of observations as follows
\begin{equation}
\mathcal{O}_1 = \left\{
\begin{array}{ll}
\bra{\psi_t} \sigma^x_1 \otimes I \ket{\psi_t}, &
\bra{\psi_t} \sigma^z_1  \otimes I \ket{\psi_t}, \\
\bra{\psi_t} I \otimes \sigma^x_2 \ket{\psi_t}, &
\bra{\psi_t} I \otimes \sigma^z_2 \ket{\psi_t}.
\end{array}
\right\},\label{eq:19}
\end{equation}
\begin{equation}
\mathcal{O}_2 = \left\{
\begin{array}{ll}
\bra{\psi_t} \sigma^x_1 \otimes I \ket{\psi_t}, &
\bra{\psi_t} \sigma^z_1  \otimes I \ket{\psi_t}, \\
\bra{\psi_t} I \otimes \sigma^x_2 \ket{\psi_t}, &
\bra{\psi_t} I \otimes \sigma^z_2 \ket{\psi_t}, \\
\bra{\psi_t} \sigma^z_1 \otimes \sigma^z_2 \ket{\psi_t}.
\end{array}
\right\},\label{eq:20}
\end{equation}
\begin{equation}
\mathcal{O}_3 = \left\{
\begin{array}{ll}
\bra{\psi_t} \sigma^x_1 \otimes I \ket{\psi_t}, &
\bra{\psi_t} \sigma^z_1  \otimes I \ket{\psi_t}, \\
\bra{\psi_t} I \otimes \sigma^x_2 \ket{\psi_t}, &
\bra{\psi_t} I \otimes \sigma^z_2 \ket{\psi_t}, \\
\bra{\psi_t} \sigma^x_1 \otimes \sigma^x_2 \ket{\psi_t}, &
\bra{\psi_t} \sigma^z_1 \otimes \sigma^z_2 \ket{\psi_t}.
\end{array}
\right\},\label{eq:21}
\end{equation}
\begin{equation}
\mathcal{O}_4 = \left\{
\begin{array}{ll}
\bra{\psi_t} \sigma^x_1 \otimes I \ket{\psi_t}, &
\bra{\psi_t} \sigma^z_1  \otimes I \ket{\psi_t}, \\
\bra{\psi_t} I \otimes \sigma^x_2 \ket{\psi_t}, &
\bra{\psi_t} I \otimes \sigma^z_2 \ket{\psi_t}, \\
\bra{\psi_t} \sigma^x_1 \otimes \sigma^x_2 \ket{\psi_t}, &
\bra{\psi_t} \sigma^z_1 \otimes \sigma^z_2 \ket{\psi_t}, \\
\bra{\psi_t} \sigma^x_1 \otimes \sigma^z_2 \ket{\psi_t}, &
\bra{\psi_t} \sigma^z_1 \otimes \sigma^x_2 \ket{\psi_t}.
\end{array}
\right\}.\label{eq:22}
\end{equation}
Here the noise level is set as $\delta=0.01$. As shown in Fig.~\ref{fig10}, the performance of the algorithm continues to improve as the number of observations increases. However, the achieved fidelity with 8 observations is worse than that with 6 observations. This may be because the reward becomes overly large with too many observations. In particular, $\sigma$ is set as 3.5 in the reward function for all the experiments. Reducing $\sigma$ may improve the performance of the algorithm with more observations.

\subsection{6-qubit QAOA}

We consider a 6-qubit example which uses QAOA to solve the combinatorial optimization problem \cite{farhi2015quantum}. The target problem-based Hamiltonian is defined as
\begin{equation}
H_0= \omega_{jk} \sum_{j=1}^{N-1} \sum_{k>j}^{N} \sigma_j^z \sigma_{k}^z \label{eq:6-1},
\end{equation}
whose ground states are the solutions to a specific combinatorial optimization problem. The control Hamiltonian is given by
\begin{equation}
H_1= -\sum_{j=1}^{N} \sigma_j^x.\label{eq:6-2}
\end{equation}
Then initial state is set as
\begin{equation}
\ket{\psi_i}=\frac{1}{\sqrt{2^N}} \otimes_{j=1}^N (\ket{0}_j+\ket{1}_j).\label{eq:6-3}
\end{equation}
The goal of QAOA is to minimize $\langle H_0\rangle$. It has been argued that even if the optimization process may not be able to reach the ground states, at least an approximate solution can be found using the quantum-classical algorithm. We let $N=6$, and randomly generate the weights $\{\omega_{jk}\}$ from a uniform distribution $U(0,1)$ for illustration.

The full set of observations are given by
\begin{equation}
\mathcal{O} = \left\{
\begin{array}{ll}
\bra{\psi_t} \sigma^z_1 \otimes \sigma^z_2 \ket{\psi_t}, &
\bra{\psi_t} \sigma^z_1 \otimes \sigma^z_3 \ket{\psi_t}, \\
\bra{\psi_t} \sigma^z_1 \otimes \sigma^z_4 \ket{\psi_t}, &
\bra{\psi_t} \sigma^z_1 \otimes \sigma^z_5 \ket{\psi_t}, \\
\bra{\psi_t} \sigma^z_1 \otimes \sigma^z_6 \ket{\psi_t}, &
\bra{\psi_t} \sigma^z_2 \otimes \sigma^z_3 \ket{\psi_t},
\\
\bra{\psi_t} \sigma^z_2 \otimes \sigma^z_4 \ket{\psi_t}, &
\bra{\psi_t} \sigma^z_2 \otimes \sigma^z_5 \ket{\psi_t}, \\
\bra{\psi_t} \sigma^z_2 \otimes \sigma^z_6 \ket{\psi_t}, &
\bra{\psi_t} \sigma^z_3 \otimes \sigma^z_4 \ket{\psi_t}, \\
\bra{\psi_t} \sigma^z_3 \otimes \sigma^z_5 \ket{\psi_t}, &
\bra{\psi_t} \sigma^z_3 \otimes \sigma^z_6 \ket{\psi_t},
\\
\bra{\psi_t} \sigma^z_4 \otimes \sigma^z_5 \ket{\psi_t}, &
\bra{\psi_t} \sigma^z_4 \otimes \sigma^z_6 \ket{\psi_t}, \\
\bra{\psi_t} \sigma^z_5 \otimes \sigma^z_6 \ket{\psi_t}.
\end{array}
\right\},
\label{eq:6-4}
\end{equation}
which can be used for the complete determination of the expectation $\langle H_0\rangle$ with respect to $\ket{\psi_t}$. In the numerical experiment, only $60\%$ of the observations are made at each time step to calculate the reward, while the other observations are just generated as random noise from $U([-1,1])$. This configuration is proposed to demonstrate the model performance under the extreme condition that $40\%$ of the observations are pure noise, which is also an instance of partially observable process as aforementioned. The real observations are made as follows
\begin{equation}
\mathcal{O} = \left\{
\begin{array}{ll}
\bra{\psi_t} \sigma^z_1 \otimes \sigma^z_2 \ket{\psi_t}, &
\bra{\psi_t} \sigma^z_2 \otimes \sigma^z_3 \ket{\psi_t}, \\
\bra{\psi_t} \sigma^z_3 \otimes \sigma^z_4 \ket{\psi_t}, &
\bra{\psi_t} \sigma^z_4 \otimes \sigma^z_5 \ket{\psi_t}, \\
\bra{\psi_t} \sigma^z_5 \otimes \sigma^z_6 \ket{\psi_t},
&
\bra{\psi_t} \sigma^z_1 \otimes \sigma^z_3 \ket{\psi_t}, \\
\bra{\psi_t} \sigma^z_2 \otimes \sigma^z_4 \ket{\psi_t}, &
\bra{\psi_t} \sigma^z_3 \otimes \sigma^z_5 \ket{\psi_t}, \\
\bra{\psi_t} \sigma^z_4 \otimes \sigma^z_6 \ket{\psi_t}.
\end{array}
\right\}.
\label{eq:6-5}
\end{equation}
Although the other noisy observations are not used to calculate the reward, they are still contained in $\vec{o}_t$ to train the agent. It should be noted that only in the last step, the reward is calculated using the full set of real observations. To be more precise, the rewards for the intermediate steps are calculated as
\begin{equation}
\begin{split}
&r_t = \sigma \bigg( \underbrace{(\omega_{12}\cdot\bra{\psi_t} \sigma^z_1\sigma^z_2 \ket{\psi_t}
+...+\omega_{46}\cdot\bra{\psi_t} \sigma^z_4 \sigma^z_6 \ket{\psi_t})}_{9 \ terms}
\\&-
\underbrace{(\omega_{12}\cdot\bra{\psi_{t+1}} \sigma^z_1 \sigma^z_2 \ket{\psi_{t+1}}+...+\omega_{46}\cdot\bra{\psi_{t+1}} \sigma^z_4 \sigma^z_6 \ket{\psi_{t+1}})}_{9 \ terms} \bigg),
\end{split}
\end{equation}
while the reward for the last step is calculated as
\begin{equation}
\begin{split}
&r_t = \sigma \bigg( \underbrace{(\omega_{12}\cdot\bra{\psi_t} \sigma^z_1 \sigma^z_2 \ket{\psi_t}+...+\omega_{56} \cdot \bra{\psi_t} \sigma^z_5 \sigma^z_6 \ket{\psi_t})}_{15 \ terms}
\\
&-
\underbrace{(\omega_{12} \cdot \bra{\psi_{t+1}} \sigma^z_1 \sigma^z_2 \ket{\psi_{t+1}}+...+\omega_{56} \cdot \bra{\psi_{t+1}} \sigma^z_5 \sigma^z_6 \ket{\psi_{t+1}})}_{15 \ terms} \bigg).
\end{split}
\end{equation}

This definition of reward function is slightly different from that of the state transfer problem, as the target state is unknown in the current case. Here, the purpose of the reward function is to encourage the decrease of $\langle H_0\rangle$ on the partially observed subsystem.

\begin{figure}[!htp]
\centering
\subfigure[]{
\includegraphics[width=8.4cm]{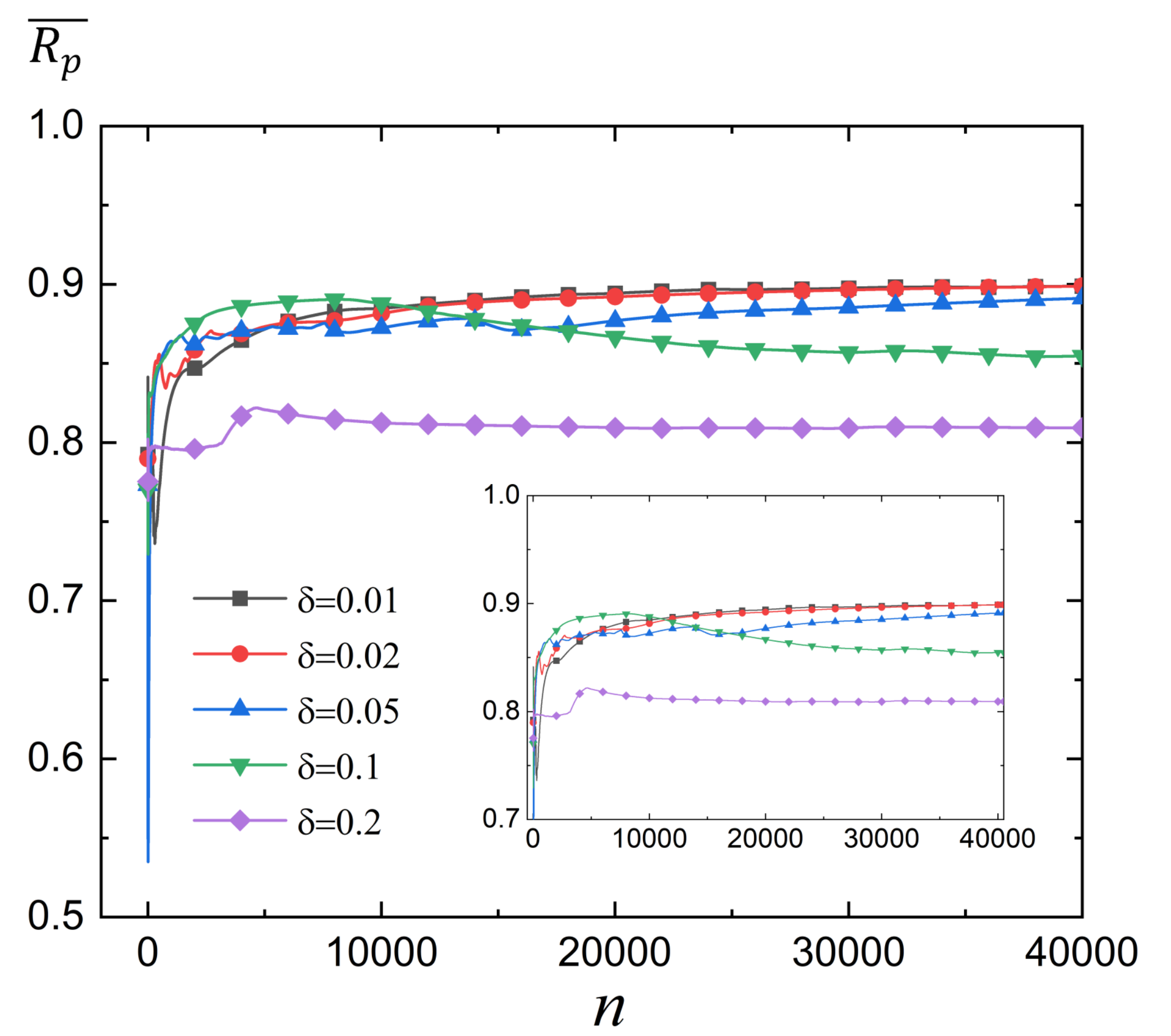}
}
\subfigure[]{
\includegraphics[width=8.4cm]{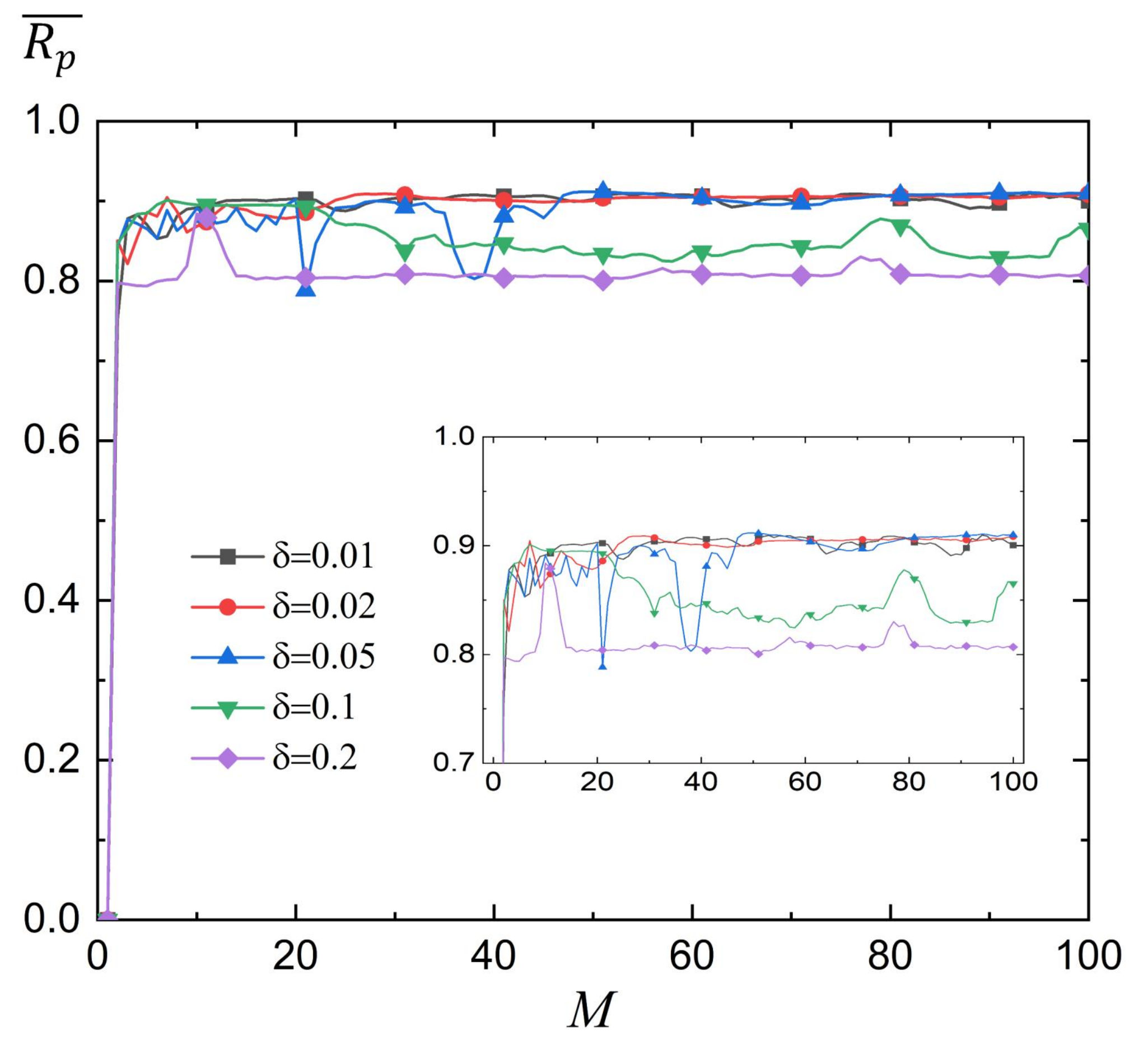}
}
\caption{The evolution of average approximation ratio $\overline{R_p}$ during training. (a) The approximation ratio averaged over the current $n$ trajectories. (b) The average approximation ratio of each epoch. Each epoch contains 400 trajectories.}
    \label{fig:6-qubit}
\end{figure}

The optimization performance is shown in Fig.~\ref{fig:6-qubit}, where we have used the following approximation ratio\cite{wauters2020reinforcement}
\begin{equation}
R_p=1-\frac{E_p(\vec{\gamma},\vec{\beta}) - E_{min}}{E_{max} - E_{min}}\label{eq:appratio}
\end{equation}
as the metric to evaluate the accuracy of the QAOA algorithm with partial observation. Here $E_{min}$ and $E_{max}$ are the lowest and highest values of $\langle H_0\rangle$. The control depth is set as $p=5$, and we let $\sigma=10$. The numerical result demonstrates that the algorithm with only $60\%$ accurate observations can still achieve a good-enough approximation ratio when $\delta \leq 0.05$. In particular, the approximation ratio can reach above $0.9$, which may yield an acceptable solution to the combinatorial optimization problem, as optimizing the approximation ratio beyond a desired value has been proven to be NP-hard with classical algorithms \cite{ZW20}.

\begin{table*}[!htp]
\centering
\caption{Comparison with the Trotterized quantum annealing algorithm.}
\label{tab3}
\setlength{\tabcolsep}{8pt}
\begin{tabular}{|c|c|c|c|}
\hline
\makecell[c]{Algorithms}    &    \makecell[c]{Accuracy}    &    \makecell[c]{Optimized Coefficients}    &    \makecell[c]{Total Duration}\\
\hline
\makecell[c]{Our method \\($\delta=0.01,\ p=5$)}     &    \makecell[c]{0.924059}    &    \makecell[c]{0.47231, 0.26109, -0.00375, \\0.00312, 0.0155, 0.00016, \\0.00896, 0.00691, 0.00445, \\0.00101}     &    \makecell[c]{0.77726}\\
\hline
\makecell[c]{Trotterized quantum annealing \\($\delta=0,\ p=5$)}   &    \makecell[c]{0.966693}&    \makecell[c]{3.27237, 3.05595, -1.36772, \\2.79921, 0.99586, -3.32931, \\-3.74419, -1.04337, -0.88244, \\-2.95511}    &    \makecell[c]{23.44553}   \\
\hline
\makecell[c]{Trotterized quantum annealing \\($\delta=0,\ p=7$)}   &    \makecell[c]{0.983772}&    \makecell[c]{2.78629, -0.37299, 0.33541, \\-2.06054, -0.93116, 0.79423, \\0.37143, -3.00828, 3.42862, \\2.18688, -1.48824, -2.03968, \\-1.7566, 0.12666}    &    \makecell[c]{21.68701}\\
\hline
\makecell[c]{Trotterized quantum annealing \\($\delta=0,\ p=9$)}   &    \makecell[c]{0.984729}&    \makecell[c]{-1.57822, 0.83495, -2.48816, \\-0.99136, -2.66433, -0.12738, \\-1.2762, 0.25729, 1.34711, \\1.75152, -0.93297, 0.98766, \\-0.76968, -0.36198, 0.79805, \\1.27066, 0.04574, -2.47126}    &    \makecell[c]{20.95452}\\
\hline
\end{tabular}
\end{table*}

We have compared our method with Trotterized quantum annealing control protocol by increasing $p$ as $p=5,7,9,11,13,15$. Part of the numerical results with optimized coefficients are shown in Table~\ref{tab3}, and the complete results are given in Supplementary Materials. The Trotterized quantum annealing algorithm is executed using PyQPanda without assuming any Hamiltonian uncertainties, and the total durations of control are initialized with the same value. Under the same control depth, the accuracy of the Trotterized quantum annealing algorithm is higher than our method. Considering that our method additionally deals with the robust control problem with Hamiltonian noise and only uses $60\%$ of full observations, such accuracy is acceptable for QAOA. It should also be noted that our strategy achieves a relatively high accuracy in the first few control steps in order to reduce the cumulative effect of Hamiltonian noise, which greatly reduces the total duration of control as compared to Trotterized quantum annealing. In practice, the control depth $p$ cannot be increased to infinity due to robustness considerations.

\section{Conclusion}

In this paper, we have proposed a quantum RL control algorithm whose reward function is calculated solely based on the partial observations. In comparison with the existing methods relying on the classical simulation and fidelity to define the reward or objective function, the proposed algorithm does not need any information other than the measurement results of partial observations, which provides an advantage for the practical implementation of control algorithms on near-term quantum devices. We show that the algorithm can accomplish high-fidelity control tasks with a significant reduction in the resource requirement for quantum measurements.

The algorithm is robust to generic noise in the control Hamiltonians and observations. In particular, the RL algorithm with partial observation can achieve similar performance under the practical level of noise when compared to the commonly-used algorithms which did not even consider the noise effect. The trained model can also work with different level of noise and different initial and target states.

The proposed framework may also find applications in the practical implementation of noisy QAOA for the demonstration of near-term quantum computing, which only requires QAOA to give an optimization result that is good enough instead of the best result. In this case, the control precision of RL method with partial observability may be high enough for achieving an acceptable level of optimization accuracy.

The optimization process of POMDP proposed in this paper may be computationally more demanding than the conventional gradient-based methods, mainly because deep RL models are more computationally demanding than conventional gradient descent. Nevertheless, since the scale of the neural network often grows linearly with the number of qubits, the computational resource required by the deep RL models will not constitute the major obstacle for scalability. The scalability issue in quantum systems comes from the exponential grow of the size of quantum state with the number of qubits. As a result, the conventional gradient-based methods still face the same scalability issues as the partially-observation method does. In certain cases, the number of partial observations only grows linearly with the number of qubits. For example, \cite{wauters2020reinforcement} has successfully carried out the simulation on the Ising chain of $128$ spins using a deep RL scheme. For Ising chain, only the adjacent pair of qubits are measured. Therefore, the computation scalability can be significantly enhanced.

Further procedures can be incorporated into the partial observation scheme to improve the robustness and reduce the measurement cost. For example, Ref.~\cite{ding2021breaking} proposed a robust quantum control protocol against systematic errors by combining Short-cuts-To-Adiabatic (STA) and deep RL methods, which has been verified in the trapped-ion system \cite{ai2022experimentally}. Moreover, weak measurements can be implemented to reduce the measurement cost more significantly \cite{borah2021measurement,ding2021quantum}, which have been successfully applied to the double-well and dissipative qubit systems, respectively. Since the state of the qubit is not destroyed but slightly perturbed via weak measurement, these protocols no longer need to record the historical data for repetitive state preparation, and thus may bring significant resource savings in practical implementations.

%-----------------------------------------------------------------

\section*{References}
\bibliography{pomdp}

\appendix

\section{Partially Observable Markov Decision Process (POMDP)}
\label{secpomdp}
The Markov Decision Process (MDP) is a sequential decision process for a fully observable, stochastic environment with a Markovian transition model and additive rewards \cite{ref.53}. By partially observable we mean that the agent only receives an observation $o_{t+1}\in\mathcal{O}$ through non-complete set of measurements. POMDP is often defined as a $6$-tuple $(\mathcal{S},\mathcal{A},\mathcal{T},\mathcal{R},\mathcal{O},\mathcal{Z})$, where $\mathcal{S},\mathcal{A},\mathcal{T},\mathcal{R}$,$\mathcal{O},\mathcal{Z}$ are state space, action space, transition function, immediate reward, observation space and conditional observation probabilities, respectively. In MDP and POMDP, the agent takes an action $a_t$ after the current observation $o_t$ at each time step $t$. Then the environment reaches the next state $s_{t+1}$ with the probability $\mathcal{T}(s_{t+1}|s_{t},\ a_t)$, and the agent receives the next observation $o_{t+1}$ with the probability $\mathcal{Z}(o_{t+1}|s_{t+1})$. The agent needs to choose actions $\{a_t\}$ in order to maximize the expected discounted reward $\mathbb{E}[\sum_{t=0}^{\infty} \gamma^t r_t]$, where $r_t$ is the immediate reward function at the current time step $t$ and $\gamma$ is the discount factor that describes the preference of the agent for current rewards over future rewards. 

\section{Reinforcement Learning (RL) and Actor-Critic Framework}
\label{secrl}
RL is an effective method to generate an optimal policy for choosing the proper actions at each time step $t$. In our algorithm, we utilize the deterministic policy $\mu$ by implementing $a_t = \mu (o_t)$. In order to learn an optimal policy, the common practice is to learn an action-value function $Q^\mu (o_t,a_t)$ which is defined as follows
\begin{equation}
Q^\mu (o_t,a_t) = \mathbb{E}[\sum_{l=0}^{\infty} \gamma^l r(o_{s+l},a_{s+l})|o_t,a_t].\label{eq:ap1}
\end{equation}
Here $Q^\mu (o_t,a_t)$ is often called the Q-value of the action $a_t$. Such a function indicates the expected reward starting from the observation $o_t$ with action $a_t$ and policy $\mu$. 

Actor-Critic \cite{konda1999actor} is a commonly-used framework of RL, in which the policy $\mu$ and the action-value function $Q$ are called actor and critic, respectively. Both the actor and the critic are implemented as neural networks. TD3 \cite{fujimoto2018addressing} is a state-of-the-art actor-critic framework which can address the problem of overestimating the Q-value. Specifically, TD3 employs two critics $Q_1$ and $Q_2$, which uses the minimum of the predicted optimal future return in state $s_{t+1}$ to bootstrap the Q-value of the current observation $o_t$ and action $a_t$ as illustrated in Fig.~\ref{fig5}. The parameters of the critic $Q_j$ are updated as follows

\begin{equation}
\theta_{Q_j} \gets \mathop{\arg \min}_{\theta_{Q_j}} \mathbb{E} [(Q_j (\vec{o}_t,a_t,h_t^l)-\hat{Q})^2]\label{eq:ap2}.
\end{equation}
The target Q-value $\hat{Q}$ is determined by 
\begin{equation}
\hat{Q}=r_t+\gamma*\mathop{\min}_{j=1,2} Q_{j} (\vec{o}_t,a_t,h_t^l),\label{eq:ap3}
\end{equation}
and the parameters of the actor are optimized as follows
\begin{equation}
\theta_{\mu} \gets \mathop{\arg \max}_{\theta_{\mu}} \mathbb{E} [Q(o_t,\mu(o_t,h^l_t),h^l_t)],\label{eq:ap4}
\end{equation}
in which the $Q$ function could be either of the two critics $Q_1$ and $Q_2$.

\end{document}